%% file: Main.tex
\documentclass[12pt,draftclsnofoot,onecolumn]{IEEEtran}
\IEEEoverridecommandlockouts

\usepackage{graphicx}
\usepackage{subfigure}
\usepackage{amsmath,amsthm,amsfonts,amssymb}
\usepackage{cite}
\usepackage{bm}
\usepackage{url}
\usepackage{array}
\usepackage{color}
\usepackage{multirow}
\usepackage{booktabs}
\usepackage[table,xcdraw]{xcolor}
\usepackage{enumerate}

\theoremstyle{plain}
\newtheorem{thm}{Theorem}

\newtheorem{prop}[thm]{Proposition}
\newtheorem{cor}[thm]{Corollary}
\newtheorem*{comm}{Comment}
\newtheorem{sty1}{Theorem}
\newtheorem{defi}[sty1]{Definition}

\newenvironment{NewProof}{{\noindent\it Proof.}}{\hfill $\blacksquare$\par}

\allowdisplaybreaks[4]

\makeatletter
\def\old@comma{,}
\catcode`\,=13
\def,{%
  \ifmmode%
    \old@comma\discretionary{}{}{}%
  \else%
    \old@comma%
  \fi%
}
\makeatother

\begin{document}
\title{Flexible Subcarrier Allocation for Interleaved Frequency Division Multiple Access}

\author{
Yulin~Shao,~\IEEEmembership{Student Member,~IEEE},
Soung~Chang~Liew,~\IEEEmembership{Fellow,~IEEE}
\thanks{Y. Shao and S. C. Liew are with the Department of Information Engineering, The Chinese University of Hong Kong, Shatin, New Territories, Hong Kong (e-mail: \{sy016, soung\}@ie.cuhk.edu.hk).}
}

\maketitle

\begin{abstract}
Interleaved Frequency Division Multiple Access (IFDMA) and Orthogonal FDMA (OFDMA) belong to a class of signal modulation and multiple-access techniques in which information of multiple users are multiplexed and carried on subcarriers within a shared spectrum.
Compared with OFDMA, IFDMA has lower Peak-to-Average Power Ratio (PAPR).
However, IFDMA poses two rigid constraints on subcarrier allocation:
1) the subcarriers occupied by a user must be evenly-spaced among the available subcarriers.
2) the number of subcarriers used by a user must be a divisor of the total number of subcarriers.
This paper investigates how to overcome these constraints to allow flexible and fine-grained subcarrier allocation in IFDMA.
Specifically, we put forth
1) a bit-reversal subcarrier allocation scheme whereby the problem of allocating evenly-spaced subcarriers is transformed to a more intuitive problem of filling contiguous bins;
2) a multi-stream IFDMA scheme whereby a user can have an arbitrary number of subcarriers.
For the synchronous scenario in which user requests arrive in a synchronous manner, we show that IFDMA can achieve the same level of flexibility and granularity as OFDMA in subcarrier allocation.
For the asynchronous scenario in which user requests arrive and depart asynchronously, we show that the blocking probability of IFDMA is only slightly worse than that of OFDMA: specifically, the gap between the blocking probabilities of IFDMA and OFDMA is only $2.56\%$ at a moderate offered load.
\end{abstract}


\section{Introduction}
Wireless networks are consuming ever more energy \cite{EE2}. For example, China Mobile operated $3.85$ million base stations (BS) in $2018$ (inclusive of $2.41$ million $4$G BS)~\cite{China1}, with an annual energy consumption of $24,470$ GWhs~\cite{China2}. This is only the energy consumption by one operating company. In the $5$G era, worldwide energy consumption of wireless networks is expected to increase by two to three times~\cite{EE3}. Energy efficiency has become a primary concern in the design and operation of future cellular, Internet of Things (IoT), and sensor networks.

Orthogonal Frequency Division Multiplexing (OFDM) accounts for a large part of the energy consumption in 5G \cite{EE3,PAPRTSP}. A major problem of OFDM signals is the large amplitude fluctuation. When passed through a nonlinear power amplifier, the large peak of OFDM signals leads to in-band distortion and out-of-band radiation \cite{PAPROFDM1,PAPROFDM2}. Thus, Highly Linear Amplifiers (HLA) are required for OFDM systems. HLA, however, incurs large power dissipation \cite{Amp}.

Interleaved Frequency Division Multiple Access (IFDMA) and Orthogonal FDMA (OFDMA) are signal modulation and multiple-access techniques that multiplex information of multiple users onto a set of subcarriers within a shared spectrum \cite{MAarticle0,MAarticle}. OFDMA is widely used in existing broadband wireless systems. It is based on OFDM and therefore also suffers from large amplitude fluctuation. Compared with OFDMA, IFDMA presents a waveform with significantly lower Peak-to-Average Power Ratio (PAPR)~\cite{PAPR1,IFDMA2}. As a result, IFDMA transmitters do not require HLA, and by avoiding the large power dissipation associated with HLA, IFDMA can be more energy saving than OFDMA~\cite{MAbook}.

The excellent PAPR performance of IFDMA comes from the combining of multi-carrier modulation with spreading~\cite{MA2,IFDMA1}. Specifically, in IFDMA, the information of a user is modulated to subcarriers that are evenly distributed across the whole spectrum. As a result, the PAPR of IFDMA is similar to that of Code Division Multiple Access (CDMA) \cite{DSCDMA}. In particular, a constant-envelope waveform (with $0$~dB PAPR) is obtained if the users adopt Phase Shift Keying (PSK) modulation and rectangular pulse shaping. Other advantages of IFDMA over OFDMA includes the ability to achieve full frequency diversity, smaller sensitivity to frequency offset, and the ability to combat frequency nulls \cite{SCBook}.

IFDMA, however, has rigid requirements on subcarrier allocation. Let us refer to a data stream multiplexed onto a set of subcarriers as an IFDMA stream. At any one time, the shared spectrum may be occupied by IFDMA streams of multiple users. An IFDMA stream is subject to the following two constraints \cite{MAarticle,IFDMA1}:
\begin{itemize}
\item {Constraint $1$}: Subcarrier allocation in IFDMA must be carefully orchestrated so that not only the subcarriers of different IFDMA streams are non-overlapping, but also that the subcarriers of each IFDMA stream are evenly spaced (e.g., subcarriers $\{0, 4, 8, ...\}$ are allocated to one IFDMA stream; subcarriers $\{1, 17, 33, ...\}$ to another IFDMA stream).
\item {Constraint $2$}: The number of subcarriers of an IFDMA stream must be a divisor of the number of the overall available subcarriers (e.g., if there are $64$ subcarriers, the number of subcarriers of an IFDMA stream must be one of $\{1, 2, 4, 8, 16, 32, 64\}$; the number of subcarriers of a stream cannot be $3$).
\end{itemize}

These two constraints lead to the following impressions:
\begin{enumerate}
\item With Constraint $1$, the subcarrier allocation of IFDMA is very inflexible, even in the synchronous scenario where all the user' requests arrive at the AP simultaneously.
\item With Constraint $1$, the subcarrier allocation of IFDMA is even more complex in the asynchronous scenario where the requests of different users can come and go in an asynchronous manner.
\item With Constraint $2$, the subcarrier allocation granularity of IFDMA is low.
\end{enumerate}

Prior works on subcarrier allocation of IFDMA, to be elaborated in Section \ref{sec:IIB}, addresses only Constraint $1$ in the synchronous scenario \cite{Allocation1,Allocation2}. In this paper, we argue that all the above three impressions dues to Constraints $1$ and $2$ are not exactly true.
In fact, the imposed rigidity of IFDMA belies nice mathematical structures that can be exploited to simplify subcarrier allocation and transceiver design
Specifically, (i) the bit-reversal mapping, and (ii) signal modulations of IFDMA streams of different sizes can be interpreted as FFTs of different scales embedded within the overall Cooley-Tukey FFT recursive decomposition scheme.
The simple IFDMA transceiver designs inspired by (ii) are treated comprehensively in our companion paper~\cite{tech2}. The main focus of this paper is on subcarrier allocation in IFDMA systems and we will exploit (i) to simplify the process.

The major contributions and conclusions of this paper are as follows:
\begin{enumerate}
\item We put forth a new bit-reversal subcarrier allocation scheme for IFDMA whereby, instead of focusing on allocating evenly-spaced subcarriers to an IFDMA
    stream, we can exploit a simpler notion of filling contiguous ``bins'' in IFDMA resource allocation. After such bin filling, a bit-reversal mapping of the contiguous bin indexes automatically yields the evenly-spaced subcarrier indexes. Based on this scheme, we design two bin-filling policies: the sort-first policy and the min-small-change policy.
\item In the synchronous scenario, we prove that both the sort-first and the min-small-change policies allow full loading in IFDMA systems. Specifically, as long as the sum total of the subcarriers requested by all users does not exceed the number of available subcarriers in the system, the two policies will be able to satisfy all the requests. Thus, IFDMA is as flexible as OFDMA in synchronous subcarrier allocation.
\item In the asynchronous scenario, we investigate the blocking properties of IFDMA. We prove that the offered load of an asynchronous IFDMA system has to be severely limited in order that the system is strictly nonblocking. That is, the allowed loading would be much lower than full loading if IFDMA were to be operated as a strictly nonblocking system. Fortunately, by allowing a small blocking probability, IFDMA can have good performance. Under moderate offered load, our min-small-change policy reduces the blocking probability of IFDMA systems by $17.21\%$ compared with the random policy in \cite{Allocation1,Allocation2}, and the gap between the blocking probabilities of IFDMA and OFDMA is only $2.56\%$.
\item We put forth a multi-stream IFDMA scheme that allows each user to be allocated multiple IFDMA streams to serve two purposes. First, Constraint $2$ is addressed with multi-stream IFDMA. The size of a request can be any positive integer smaller than or equal to the total number of available subcarriers. Second, the multi-stream IFDMA systems are strictly nonblocking even under full loading. Overall, multi-stream IFDMA is on an equal footing with OFDMA as far as the flexibility and granularity of subcarrier allocation are concerned. Yet, multi-stream IFDMA has better PAPR performance than OFDMA.
\end{enumerate}

\section{Subcarrier Allocation in IFDMA}\label{sec:II}
\input{SecII.tex}

\section{Overview of Our Key Results}\label{sec:III}
\input{SecIII.tex}

\section{Synchronous Subcarrier Allocation}\label{sec:IV}
\input{SecIV.tex}

\section{Asynchronous Subcarrier Allocation}\label{sec:V}
\input{SecV.tex}

\section{Conclusion}\label{sec:Conclusion}
Interleaved Frequency Division Multiple Access (IFDMA) is a multiplexing and signal modulation technique for multi-user systems. It has attracted great interest because of its excellent Peak-to-Average Power Ratio (PAPR).

However, the strict constraints of IFDMA subcarrier mapping (i.e., the evenly-spaced-sub\-carrier constraint and the divisor-of-$M$ constraint) led to the following impressions:
1)	The subcarrier allocation of IFDMA is not only complicated, but also very inflexible, especially in the asynchronous scenario where the requests of different users can come and go in an asynchronous manner.
2)	The granularity of IFDMA subcarrier allocation is low in that the number of subcarriers assigned to an IFDMA data stream must be a divisor of the total number of available subcarriers.

The results of this paper indicate otherwise. {it With the bit-reversal subcarrier allocation and the multi-stream IFDMA schemes proposed in this paper, flexible and fine-grained resource allocation is possible for IFDMA in both synchronous and asynchronous scenarios.} Specifically,
\begin{enumerate}
\item in the synchronous scenario, all requests can be satisfied provided that the sum total of the requested subcarriers does not exceed the number of available subcarriers in the system;
\item in the asynchronous scenario, the blocking probability of IFDMA is only slightly worse than that of OFDMA: specifically, the gap between the blocking probabilities of IFDMA and OFDMA is only $2.56\%$ at a moderate offered load.
\end{enumerate}

In essence, major shortcomings of IFDMA can be overcome with our proposed schemes. With the shortcomings thus removed, given its excellent PAPR performance (hence excellent energy efficiency), IFDMA is a serious green technology for future wireless communications networks.

\appendices
\section{Digit-reversal Subcarrier Allocation for Composite-$M$ IFDMA}\label{sec:AppA}
\input{AppendixA.tex}

\section{Proof of Proposition~\ref{thm:8}}\label{sec:AppB}
\input{AppendixB.tex}
\section{Proof of Proposition~\ref{thm:9}}\label{sec:AppC}
\input{AppendixC.tex}

\bibliographystyle{IEEEtran}
\bibliography{References}

\end{document}

%% file: SecII.tex
\subsection{IFDMA}
We consider an uplink multiple access system where multiple users transmit signals to a common access point (AP) using orthogonal subcarriers. When operated with IFDMA, each user requests a set of subcarriers according to the rate requirement, and constructs an IFDMA stream to transmit the signal.

Let $M$ be the total number of available subcarriers, and $N$ be the number of subcarriers requested by a user. In order for the user to be able to construct an IFDMA stream, the subcarrier allocation is subject to two constraints:
\begin{itemize}
\item {Constraint $1$}: The $N$ subcarriers allocated to a user must be evenly spaced among the $M$ subcarriers. That is, the distance between the indexes of two adjacently allocated subcarriers must be $M/N$.
\item {Constraint $2$}: $N$ must be a divisor of $M$ so that $M/N$ is an integer.
\end{itemize}

If we index the $M$ subcarriers by $[M]=\{0,1,\cdots,M-1\}$, then the indexes of the $N$ subcarriers allocated to the user are
\begin{eqnarray}\label{eq:II1}
\left\{ d+i\frac{M}{N}:i=[N]  \right\},
\end{eqnarray}
where $d\in [M/N]$ is a constant frequency shift.

\begin{figure}[t]
  \centering
  \includegraphics[width=0.8\columnwidth]{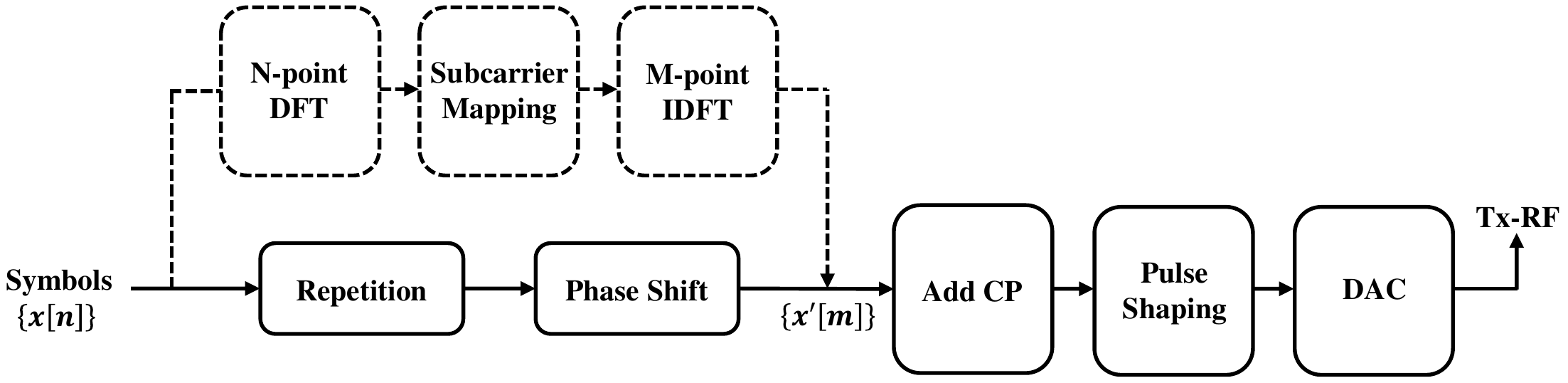}\\
  \caption{Construction of an IFDMA stream of size $N$ at the transmitter.}
\label{fig:1}
\end{figure}

Given the $N$ allocated subcarriers, Fig.~\ref{fig:1} shows two possible realizations for the construction of an IFDMA stream at the transmitter. The first is a realization in the frequency domain (the dashed modules), and the second is a realization in the time domain (the solid modules).

Let us first focus on the frequency-domain realization. As shown, the symbols to be transmitted are grouped into blocks of size $N$, that is, $\{x[n]:n=[N]\}$. Each block of time-domain symbols is transformed to the frequency domain by an $N$-point DFT, and is then mapped to the $N$ allocated subcarriers according to \eqref{eq:II1}, with the $M-N$ unoccupied subcarriers filled with $0$. After that, the transmitter performs an $M$-point IDFT to transform the data on $M$ subcarriers back to the time-domain, and obtains an IFDMA stream $\{x'[\ell]:\ell=[M]\}$. It is easy to verify that \cite{MAarticle}
\begin{eqnarray}\label{eq:II2}
x'[\ell]=\frac{N}{M}e^{j\frac{2\pi \ell}{M}d} x[\ell~\textup{mod}~N],~~\ell=[M].
\end{eqnarray}
In particular, if $d=0$ (i.e., the allocated subcarriers are $\{i\frac{M}{N}:i=[N]\}$), the IFDMA stream is simply $x'[\ell]=\frac{N}{M}x[\ell~\textup{mod}~N]$, $\ell=[M]$.

Eq.~\eqref{eq:II2} gives the simpler and more direct time-domain realization of the IFDMA stream. That is, $\{x'[\ell]\}$ can be directly constructed from $\{x[n]\}$ by repetition and frequency upshifting, hence bypassing the $N$-point DFT, subcarrier mapping/multiplexing and the $M$-point IDFT.

\subsection{Prior Solutions to Subcarrier Allocation}\label{sec:IIB}
Most of the nice attributes of IFDMA can be understood from its simple time-domain representation in~\eqref{eq:II2}, e.g., the simple transmitter design in uplink and the superior PAPR property compared with OFDMA.

Despite the advantages, a common view on IFDMA is that the subcarrier allocation, as well as the multiplexing/demultiplexing of IFDMA streams, are quite inflexible due to the above two constraints. Specifically,
\begin{itemize}
\item Constraint $1$ requires the subcarriers allocated to each user to be evenly spaced among the M subcarriers. This is even more challenging in asynchronous systems (defined below) where the requests from different users can come and go in an asynchronous manner.
\item Constraint $2$ prevents the users from requesting an arbitrary number of subcarriers, which is possible in OFDMA. IFDMA thus has coarse granularity in resource allocation.
\end{itemize}

\begin{defi}[Synchronous and asynchronous subcarrier allocation]
In synchronous subcarrier allocation, all the users' requests arrive at and depart from the AP simultaneously. A new batch of requests will not arrive until all the earlier requests have left the system. In asynchronous subcarrier allocation, the users' requests arrive at the AP asynchronously, and the service times of different requests are non-uniform. Thus, a newly arriving request may see some subcarriers already occupied by existing requests.
\end{defi}

Prior works on subcarrier allocation address only Constraint $1$ in synchronous systems. The general idea is to construct a tree structure, and associate the vertices of the tree with eligible sets of subcarriers that can be allocated to users.

\begin{figure}[t]
  \centering
  \includegraphics[width=0.65\columnwidth]{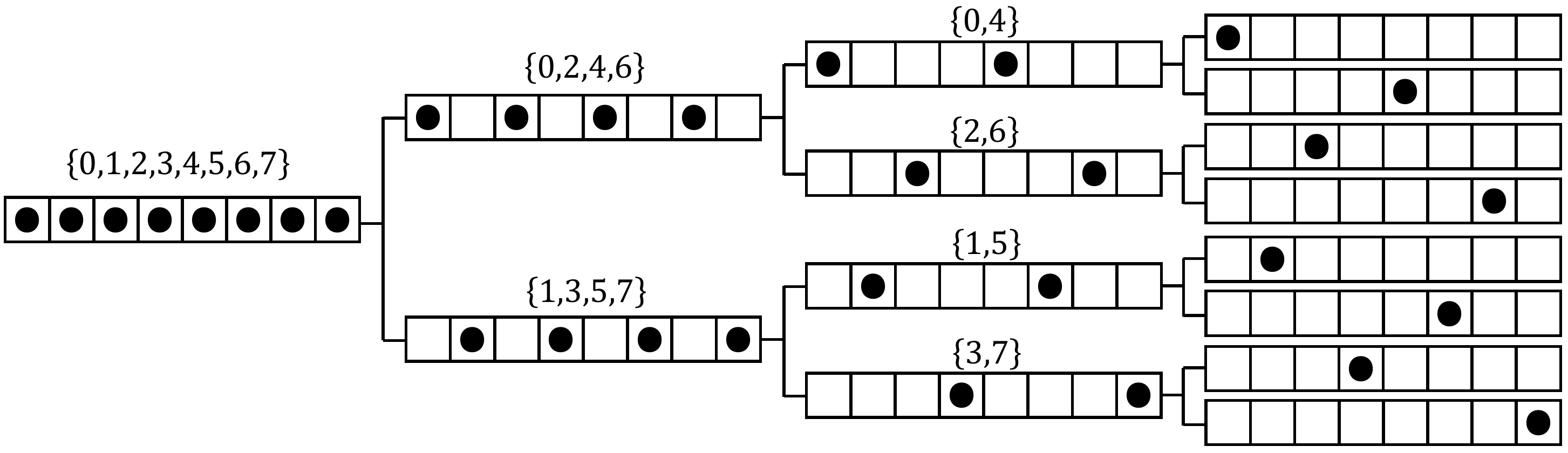}\\
  \caption{A tree structure for synchronous subcarrier allocation wherein $M = 8$. Each vertex of the tree is associated with a set of subcarriers (represented by black points) that can be assigned to users.}
\label{fig:2}
\end{figure}

In~\cite{Allocation1}, the author assumed $M$ is a power of $2$, and constructed a tree similar to the one used for generating Orthogonal Variable Spreading Factor (OVSF) codes. An example of the tree structure is given in Fig.~\ref{fig:2}.

As can be seen, each vertex of the tree is associated with a set of subcarriers (represented by black points) that can be assigned to users. Given a number of synchronous requests, the AP performs the following procedure to allocate subcarriers
\begin{enumerate}[Step i --]
\item Sort the requests according to their sizes in descending order.
\item For the request with the largest size, randomly choose a vertex on the tree whose associated set has the same size with the request. Assign the subcarriers in the associated set to the request.
\item Eliminate the vertex chosen in step ii and all its descendent vertex.
\item Repeat steps ii and iii until all requests are satisfied.
\end{enumerate}

\begin{figure}[t]
  \centering
  \includegraphics[width=0.45\columnwidth]{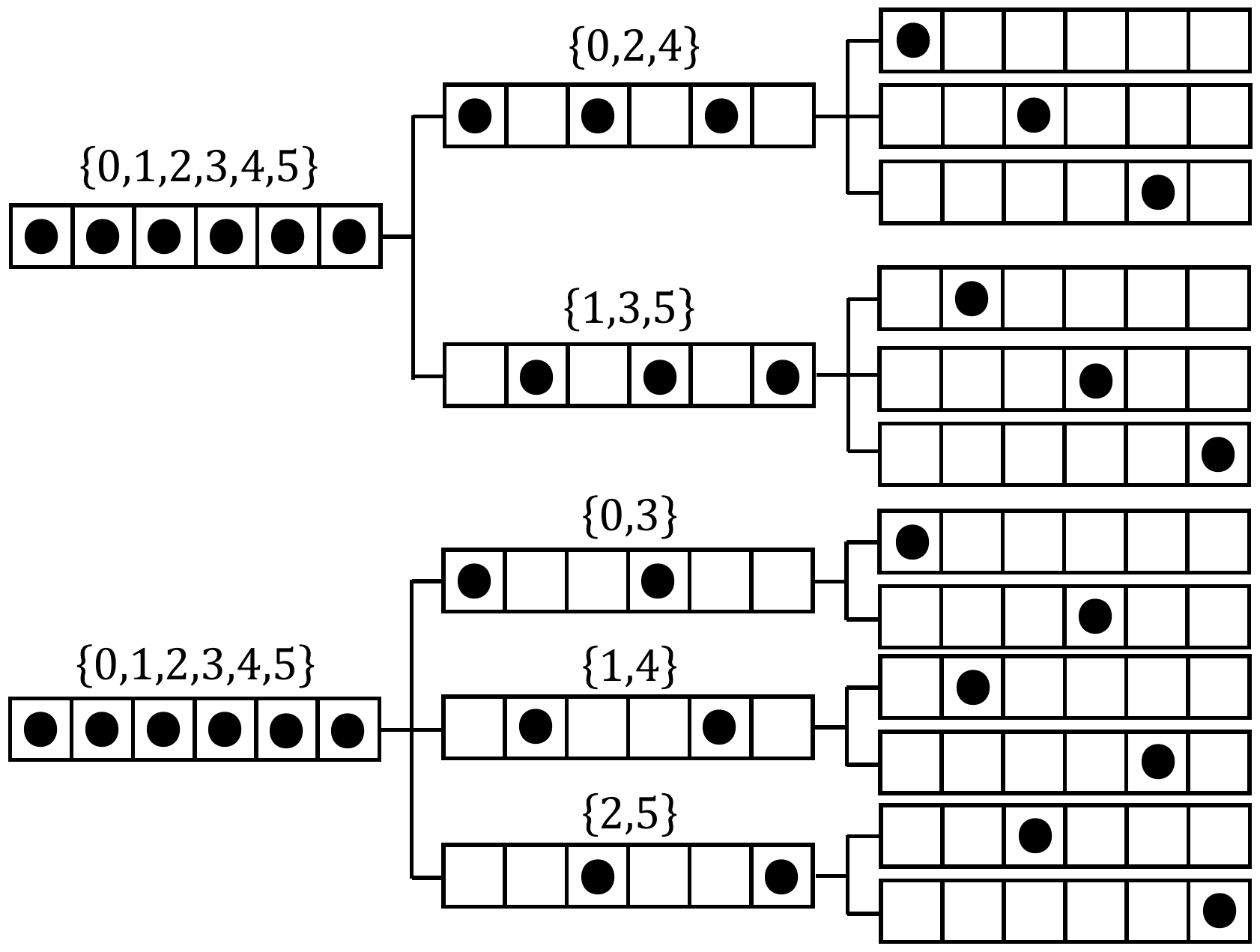}\\
  \caption{The two tree structures for subcarrier allocation when $M = 6$.}
\label{fig:3}
\end{figure}

In~\cite{Allocation2}, the authors studied the synchronous subcarrier allocation problem for more general composite $M$. Unlike the case of power-of-$2$ $M$ where the binary tree is well-determined, the tree structure for general composite $M$ has many variations. For example, Fig.~\ref{fig:3} presents the two variations when $M = 6$.

To summarize,
\begin{enumerate}
\item Prior solutions address only the synchronous subcarrier allocation. Their idea is to build a tree structure so that the vertices of the tree can be mapped to different sets of evenly-spaced subcarriers of various sizes.
\item Prior solutions cannot be generalized to asynchronous subcarrier allocation because sorting the requests is no longer possible when request arrivals are asynchronous. Without sorting, the random vertex-selection policy leads to a large blocking probability of incoming requests, as will be shown in this paper.
\item Prior solutions still suffer from Constraint $2$. $N$ must be a divisor of $M$, but not any integers, limiting the granularity of the allocated resources.
\end{enumerate}

%% file: SecIII.tex
This paper makes advances on all the three problems faced by prior works, as elaborated below:

1) For synchronous subcarrier allocation, we put forth a bit-reversal\footnote{Bit reversal is for power-of-$2$ $M$. For general composite $M$, instead of bit-reversal, we use digit-reversal, as detailed in Appendix~\ref{sec:AppA}. To ease exposition, we assume the total number of subcarriers $M$ is a power of $2$ in the main body of this paper.} subcarrier allocation scheme where no tree construction is required. Given this scheme, we prove that resource allocation of IFDMA is on an equal footing with OFDMA in that ``full loading'' is possible. That is, provided constraints $1$ and $2$ are fulfilled by all user requests, as long as the total number of subcarriers requested by the users is no more than $M$, all the requests can be fulfilled.

An important finding which form the basis for our bit-reversal subcarrier allocation scheme is that, the indexes of the evenly-spaced subcarriers is contiguous in a bit-reversal permutation \cite{bitReversal}.
For example, the bit-reversal permutation for $M = 8$ is $(0,4,2,6,1,5,3,7)$.
For IFDMA, a user requesting two subcarriers can be allocated subcarriers indexed by $\{0,4\}$, $\{1,5\}$, $\{2,6\}$, or $\{3,7\}$; a user requesting four subcarriers can be allocated subcarriers indexed by $\{0,2,4,6\}$ or $\{1,3,5,7\}$.
Suppose that each subcarrier is put into a bin with the bin index being the bit reversal of the subcarrier index. Then the subcarriers allocated to a user will occupy bins with contiguous bin indexes. For example, subcarriers $\{1,3,5,7\}$ will occupy bins $\{4,5,6,7\}$.

Thus, instead of subcarrier allocation, we can consider an equivalent problem of bin allocation; and instead of evenly-spaced subcarriers, we can now have the simpler mental picture of contiguous bins when performing resource allocation. As a result, the problem of allocating evenly-spaced subcarriers is transformed to the more intuitive problem of filling contiguous bins, and no tree construction is needed for synchronous IFDMA systems. This model also comes in handy when we consider asynchronous subcarrier allocation.

2) For asynchronous subcarrier allocation, we put forth a new bin-filling policy, the ``min-small-change policy'' (to be elaborated in Section \ref{sec:IV}), to avoid fragmentation of contiguous bins, hence reducing blocking probability. Compared with the random bin-filling policy, our min-small-change policy reduces the blocking probability of IFDMA systems by $17.21\%$ under moderate offered load in a system where the arrivals are a Poisson process \cite{StochasticBook} and the holding times of accepted requests are exponentially distributed.

\begin{figure}[t]
  \centering
  \includegraphics[width=0.8\columnwidth]{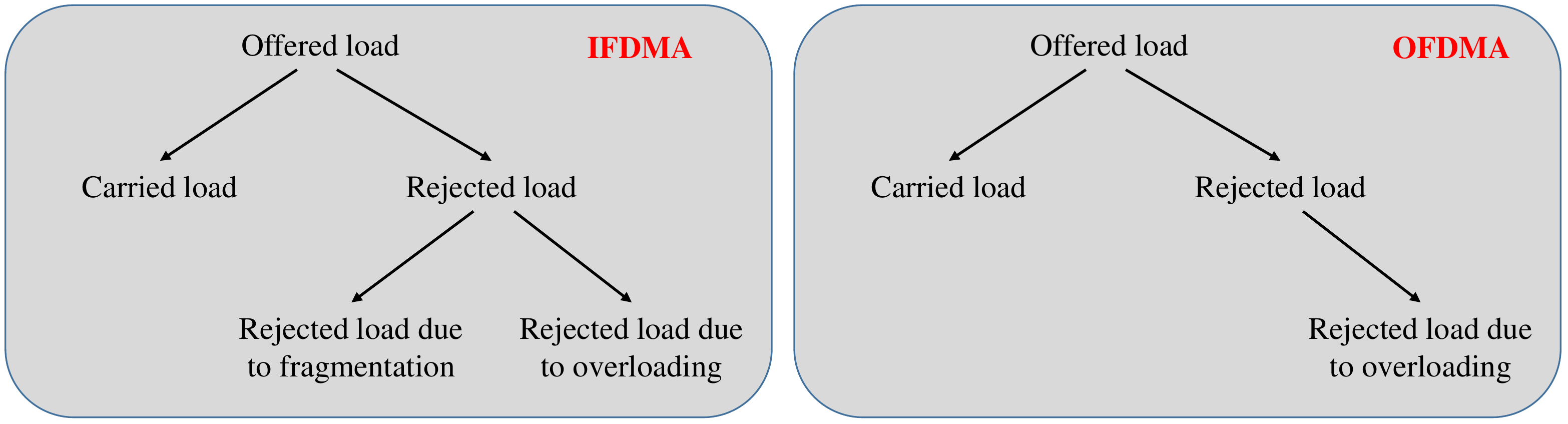}\\
  \caption{The partition of the offered load in asynchronous IFDMA and OFDMA systems.}
\label{fig:4}
\end{figure}

Under random arrivals, an incoming request can be rejected due to instantaneous system overload (i.e., not enough subcarriers left to fulfil the number of subcarriers requested). An OFDMA system can suffer from blocking due to instantaneous overload only, as illustrated in Fig.~\ref{fig:4}. For an IFDMA system, besides blocking due to overload, an incoming request can also be rejected due to fragmentation of bins (i.e., the number of vacant bins is larger than the size of the request, but a set of contiguous bins is not available to fulfill the request).

An important metric to measure the quality of a bin-filling policy in asynchronous IFDMA systems is the ability to reduce the amount of rejected load due to fragmentation. Our min-small-change bin-filling policy aims to minimize the amount of such rejections.

However, even with our min-small-change policy, the asynchronous IFDMA systems still have higher blocking probability than asynchronous OFDMA systems because the amount of rejections caused by fragmentation cannot be reduced to $0$. In this context, this paper further proposes a multi-stream IFDMA scheme, as described below.

3) We put forth a new multi-stream IFDMA scheme to serve two purposes. First, it enables us to remove Constraint $2$. The request size $N$ can be any positive integer smaller than or equal to $M$ in multi-stream IFDMA systems. Second, the multi-stream IFDMA can achieve the same level of flexibility with OFDMA even in asynchronous systems (i.e., a request can be rejected only due to system overload, but not due to bin fragmentation).

\begin{defi}[Multi-stream IFDMA]\label{def:2}
Let $M$ be the total number of subcarriers. Suppose a user request for $N$ subcarriers, where $N$ can be any positive integer smaller than or equal to $M$. A multi-stream IFDMA system allows the user to construct multiple IFDMA streams, as opposed to only one stream in traditional IFDMA systems. Specifically, the AP will partition the request to $Q$ sub-requests, and the size of each sub-request is a divisor of $M$. That is, $N=\sum_{i=1}^{Q}N_i$ and each $N_i$ is a divisor of $M$. The AP then allocates subcarriers to each sub-request following the evenly-spaced-subcarrier rule. The user will then construct $Q$ IFDMA streams and transmit the sum of the $Q$ streams.
\end{defi}

For example, suppose that $M =16$ and $N = 6$. In a conventional IFDMA system, the user has to request for $8$ instead of $6$ subcarriers to fulfill its rate requirement because $6$ is not a divisor of $16$, causing bandwidth wastage. However, in multi-stream IFDMA systems, the user can partition $6$ into $4 + 2$, and construct two IFDMA streams, one of size $4$ and the other of size $2$. The transmitted signal of this user is then the sum of the two IFDMA streams. Note that the partition is not unique. In the extreme case, the user can construct $6$ streams, each of size $1$.

With the multi-stream IFDMA, the amount of rejections caused by fragmentation can be minimized to $0$ in asynchronous systems. This can be easily understood from the definition because we can always partition the request to smaller sub-requests. In the extreme case, multi-stream IFDMA becomes OFDMA if we partition any incoming request to multiple sub-requests of size $1$.

Despite the advantages in subcarrier allocation, multi-stream IFDMA also raises two new issues:
\begin{enumerate}[a)]
\item The multiplexing/demultiplexing of multiple IFDMA streams at both transmitter and receiver may result in more complex transceiver design.
\item The perfect PAPR property of conventional IFDMA is no longer retained, because the sum of IFDMA streams causes fluctuations in the amplitude of the transmitted signal.
\end{enumerate}

These two concerns are addressed in our companion paper on IFDMA transceiver design~\cite{tech2}. A quick summary of the content and key results of~\cite{tech2} are as follows:
\begin{enumerate}[i)]
\item In~\cite{tech2}, we devise a class of new transceiver designs for IFDMA systems that are significantly less complex than conventional IFDMA transceiver. The simple new designs are founded on a key observation that multiplexing and demultiplexing of IFDMA data streams of different sizes are coincident with the IFFTs and FFTs of different sizes embedded within the Cooley-Tukey recursive FFT~\cite{FFT1} decomposition scheme. Our new transceivers are unified designs in that they can be used in conventional IFDMA as well as multi-stream IFDMA systems. In other words, our multi-stream IFDMA transceivers do not incur extra complexity and they have the same low complexity as our single-stream IFDMA transceivers.
\item In~\cite{tech2}, we study the PAPR performance of multi-stream IFDMA benchmarked against Localized FDMA (LFDMA) and OFDMA. It is shown that the {\it minimal partition} (the partition results in the minimum number of sub-requests) yields the best PAPR performance. Overall, the multi-stream IFDMA with minimal partition is better than LFDMA, and significantly better than OFDMA in terms of PAPR performance.In other words, the PAPR of multi-stream IFDMA is not as good as that of single-stream IFDMA; but it is still better those of other schemes.
\end{enumerate}

The current paper focuses on only the resource allocation issues arising from IFDMA and multi-stream IFDMA. The reader is referred to~\cite{tech2} on how issues a) and b) are addressed.

%% file: SecIV.tex
This section focuses on the subcarrier allocation problem in synchronous IFDMA systems. We show that IFDMA is as flexible as OFDMA and LFDMA in terms of synchronous subcarrier allocation. To ease exposition, this section assumes
a) the total number of available subcarriers, $M$, is a power of $2$;
b) the number of subcarriers requested by each user, $N$, is a divisor of $M$ (hence is also a power of $2$). The results obtained in this section can be well-extended to general composite $M$ (as detailed in Appendix~\ref{sec:AppA}), and general $N$ which may not be a divisor of $M$ (with our multi-stream IFDMA scheme).

\subsection{Bit-Reversal Subcarrier Allocation Scheme}\label{sec:IVA}
To begin with, we put forth a new bit-reversal subcarrier allocation scheme.
Instead of allocating evenly-spaced subcarriers to users, we consider the problem of filling contiguous bins with users' requests. Algo. 1 presents a basic version of the bit-reversal scheme with a sort-first bin-filling policy.

\vspace{0.4cm}
\noindent \textbf{Algo 1: The bit-reversal subcarrier allocation scheme.}

\noindent Suppose that we have $M=2^m$ bins (instead of subcarriers) labelled with $k=0,1,\cdots,2^m-1$.
\begin{itemize}
\item Sorting. We sort the users according to the number of requested subcarriers in descending order.
\item Bin Allocation. We allocate bins to the sorted users in the order of bin $0$ to bin $2^m-1$. That is, each user who requests $2^i$ subcarriers is allocated the next $2^i$ unallocated bins. Users with larger request size is considered first.
\item Bit-reversal mapping. We perform a bit-reversal mapping from bins to subcarriers. That is, if a bin with binary index $b_{m-1}b_{m-2}\cdots b_{1}b_{0}$ is allocated to a user, then the subcarrier with binary index $b_{0}b_{1}\cdots b_{m-2}b_{m-1}$ will be allocated to this user.
\end{itemize}

\vspace{0.4cm}
Table~\ref{tab:1} gives an example of the bit-reversal subcarrier allocation scheme. In this example, we assume $M=8$ ($m=3$), and three users $A$, $B$ and $C$ requests for $1$, $4$ and $2$ subcarriers, respectively. To use the bit-reversal subcarrier allocation scheme, we first sort the users according to their requests. The sorted users are $B$ ($4$ subcarriers), $C$ ($2$ subcarriers) and $A$ ($1$ subcarrier). Following the second step, we allocate bins $\{0, 1, 2, 3\}$ to user $B$, bins $\{4, 5\}$ to user $C$, and then bins $\{6\}$ to user $A$. After bit-reversal mapping, users $A$, $B$ and $C$ finally get subcarriers indexed by $\{0, 2, 4, 6\}$, $\{1, 5\}$ and $\{3\}$, respectively.

\begin{table}[t]
\caption{An example of the bit-reversal subcarrier allocation.}
\center
\begin{tabular}{cccc}
\toprule
\rowcolor[HTML]{EFEFEF}
\textbf{Bins}          & \textbf{Binary index of bins} & \textbf{Bit-reversal} & \textbf{Subcarriers} \\
\midrule
0 & $000$ & $000$ & $0$ \\
1 & $001$ & $100$ & $4$ \\
2 & $010$ & $010$ & $2$ \\
3 & $011$ & $110$ & $6$ \\
4 & $100$ & $001$ & $1$ \\
5 & $101$ & $101$ & $5$ \\
6 & $110$ & $011$ & $3$ \\
7 & $111$ & $111$ & $7$ \\
\bottomrule
\end{tabular}
\label{tab:1}
\end{table}

Essentially, our bit-reversal subcarrier allocation scheme reveals that the underlying relationship between evenly-spaced subcarriers and contiguous bins is a bit-reversal mapping. In this context, we can focus on filling the contiguous bins instead of evenly-spaces subcarriers, and no tree construction is needed as in~\cite{Allocation1,Allocation2}. In fact, if we go back to tree in Fig.~\ref{fig:2}, it can be found that the subcarrier locations in the leaf vertices form a bit-reversal permutation. The tree structure and the bit-reversal mapping are closely related.

Given the bit-reversal scheme, we prove in Theorem~\ref{thm:1} below that IFDMA is as flexible as OFDMA in terms of synchronous subcarrier allocation. In particular, provided that the total number of subcarriers requested by all the users does not exceed $M$, all the requests can be fulfilled, and full loading is possible for IFDMA systems. The validity of bit-reversal mapping is also proved within the proof of Theorem~\ref{thm:1} (i.e., bit reversing the indexes of allocated contiguous bins give rise to subcarrier indexes that are evenly spaced).

\begin{thm}[Full loading is possible in IFDMA]\label{thm:1}
Let $a_n$ be the number of users requesting for $2^n$ subcarriers, and $M=2^m$. If the inequality in~\eqref{eq:IV1} below is satisfied, we can use the bit-reversal subcarrier allocation scheme to assign subcarriers to users such that the spacing between the subcarriers assign to a user with request size $2^n$ is $2^{m-n}$ (i.e., the evenly-spaced-subcarrier constraint is satisfied).
\begin{eqnarray}\label{eq:IV1}
\sum_{n=0}^{m}a_n 2^n \leq 2^m.
\end{eqnarray}
\end{thm}

\begin{NewProof}
We prove Theorem~\ref{thm:1} by induction. First, the theorem can be easily verified to be valid for $m=1$ or $2$. We prove below that if the theorem is valid for any $m\geq 2$, it is also valid for $m+1$.

A bit-reversal permutation is a permutation of a sequence of $M=2^m$ items. It is defined by indexing the elements of the sequence by the numbers from $0$ to $M-1$ and then reversing the binary representations of each of these numbers\cite{bitReversal}. Each item is then mapped to the new position given by this reversed value. For example, a length-$4$ permutation is $(0,2,1,3)$, and a length-$8$ permutation is $(0,4,2,6,1,5,3,7)$.

An observation is that, the length-$(m+1)$ bit-reversal permutation can be generated from the length-$m$ bit-reversal permutation by doubling the numbers in the length-$m$ permutation and concatenate the resulting sequence with the same sequence with each value increased by one. For example, doubling the length-$4$ permutation $(0,2,1,3)$ gives $(0,4,2,6)$; adding one to each value in $(0,4,2,6)$ gives $(1,5,3,7)$; and concatenating these two sequences gives the length-$8$ permutation $(0,4,2,6,1,5,3,7)$.

Consider bin allocation associated with the second step of our bit-reversal allocation algorithm. We have two possible cases as far as the set $\{a_n:n=1,2,\cdots,m+1\}$ is concerned: Case I -- $a_{m+1}=1$; Case II -- $a_{m+1}=0$. For case I, there is a single user wanting all the $2^{m+1}$ subcarriers. The theorem is trivially true for this case.

For case II, we divide the bins into two subsets: subset $i$ consists of bins $\{0,1,\cdots,2^m-1\}$ and subset $ii$ consists of bins $\{2^m,2^m+1,\cdots,2^{m+1}-1\}$.
Since $a_{m+1}=0$, and that all the users want only a multiple of $2^n$ ($n=0,1,2,\cdots,m$) bins, the bins allocated to a user must all belong to one of the two subsets according to our bin allocation algorithm in the second step (i.e., we cannot have some bins of a user belonging to subset $i$ while other bins of the same user belonging to subset $ii$).

Let us first consider the users being allocated bins in subset $i$. If we were to consider the length-$m$ bit reversal permutation for this group of users, the subcarriers allocated to them would satisfy the length-$m$ problem by the assumption of our inductive argument. Specifically, if a user wants $2^n$ subcarriers, $0\leq n\leq m$, then the spacing between subcarriers allocated would be $2^{m-n}$ by assumption. When we migrate to the length-$(m+1)$ problem, the indexes of the subcarriers allocated to these users would be doubled. Specifically, the new spacing would be $2\times 2^{m-n}=2^{m+1-n}$, thus satisfying the new IFDMA constraint.

We next consider the users being allocated bins in subset $ii$. If we were to index these bins by $\{0,1,\cdots,2^m-1\}$ rather than $\{2^m,2^m+1,\cdots,2^{m+1}-1\}$, we see that, by the same reasoning as in the previous paragraph, these users would also satisfy the new IFDMA constraint for the length-$(m+1)$ problem. The rest is adding one to the indexes of the allocated subcarriers so that these subcarriers are distinct from the subcarriers allocated to users of the first group in the previous paragraph. Adding one to all subcarrier indexes preserves the spacing between subcarriers. Thus, the IFDMA constraint for the length-$(m+1)$ problem remains satisfied.
\end{NewProof}

\subsection{Bin-filling Policies}
With the bit-reversal scheme, the problem of allocating evenly-spaced subcarriers is transformed to a more intuitive problem of filling contiguous bins. Algo. $1$ in Section~\ref{sec:IVA} is a bin-filling policy that satisfies the subcarrier allocation constraints. There could be other bin-filling policies that also satisfy the subcarrier allocation constraints. In this subsection, we give another policy that does so without the need for the sorting step in Algo $1$. By getting rid of the need for sorting, this policy also suggests what we should do for asynchronous subcarrier allocation where user requests do not arrive together (in which case sorting the requests before allocating subcarriers to them is not possible). This part considers synchronous subcarrier allocation with this new policy, and asynchronous subcarrier allocation will be considered in Section~\ref{sec:V}.

To satisfy the IFDMA constraint, an eligible bin-filling policy must allocate an ``unoccupied fillable bin subset'' to each user. An unoccupied fillable bin subset is basically a subset of as-yet unallocated bins with contiguous indexes that can be used to fulfill a new request, defined below:

\begin{defi}[Fillable bin subset]
A fillable bin subset of size $N$, denoted by $[iN,iN+N-1]$, is a subset of contiguous bins with labels $\{iN,iN+1,\cdots,iN+N-1\}$ such that a request of size $N$ can be allocated the bins in the subset without violating the IFDMA constraint. In particular, given a request of size $N$, there are $M/N$ ``potential fillable bin subsets'' that can be used to fulfill it, i.e., $i=0,1,2,\cdots,M/N-1$. A potential fillable bin subset can be occupied or unoccupied. For the ``occupied fillable bin subset'', at least one bin in the subset is already occupied by another request. For the ``unoccupied fillable bin subset'', all the $N$ bins in the subset are as yet unallocated and therefore can be used to fulfill a request of size $N$.
\end{defi}

\noindent\textbf{The min-small-change bin-filling policy} --
The idea of the min-small-change policy is to partition the unallocated bins into as few fillable bin subsets as possible to prevent fragmentation. With this notion, initially, without any request, we have a single unoccupied fillable bin subset, $[0,M-1]$. Suppose that the first request presented to us is a request for two bins. We can give it bins $\{0,1\}$. Now, the remaining unoccupied fillable bin subsets are $\{2,3\}$, $\{4,5,6,7\}$, $\{8, 9, 10, 11, 12, 13, 14, 15\}$, $\cdots$. If the next request presented to us is also for two bins, we should give it the subset $\{2,3\}$ rather than fragmenting other subsets for the allocation. If the request is for four bins, we should give it $\{4, 5, 6, 7\}$. The idea is analogous to keeping as little small change as possible when we spend our money. Large fillable bin subsets correspond to bills of large denominations, and small fillable bin subsets correspond to bills of small denominations. That is where the name ``min-small-change'' is coined from.

\begin{thm}[The min-small-change policy enables fully loaded IFDMA systems]\label{thm:2}
Let $M=2^m$ and $N=2^n$ where $n\leq m$. Suppose that a set of requests are presented to us in a random order in synchronous IFDMA systems. When using the min-small-change policy to fill the bins for the bit-reversal subcarrier allocation scheme, a new request for $N$ subcarriers is guaranteed to be granted provided that \eqref{eq:IV1} is satisfied (i.e., the total number of requested subcarriers in the system does not exceed $M$).
\end{thm}

\begin{NewProof}
We prove Theorem~\ref{thm:2} by contradiction. Suppose at some point during the allocation process, we are faced with a situation where the next request is for $N$ bins but we cannot find an unoccupied fillable bin subset with $N=2^n$ or more bins. If \eqref{eq:IV1} is satisfied, there must be multiple smaller unoccupied fillable bin subsets such that together they have $N$ bins. These smaller bin subsets must each have $2^i$ bins for some $i<n$. In other words, we have
\begin{eqnarray}\label{eq:IV2}
2^n=\sum_{i=0}^{n-1}\psi_i2^i,
\end{eqnarray}
where $\psi_i$ is the number of unoccupied fillable bin subsets of the size $2^i$ in the system.

On the other hand, a key observation from the bin allocation process of the min-small-change policy is that at any one time as we allocate bins to successive requests, we cannot have two unoccupied fillable bin subsets of the same size thanks to the min-small-change policy. That is, for each $i$, there can be at most one such bin subset. This means $\psi_i$ can only be either $1$ or $0$. However, in this case, we must have
\begin{eqnarray}\label{eq:IV3}
2^n>\sum_{i=0}^{n-1}\psi_i 2^i,
\end{eqnarray}
because the LHS is larger than the RHS even if $\psi_i=1$ for all $i<n$.

Eq.~\eqref{eq:IV2} contradicts \eqref{eq:IV3}. Thus, the next request for $N$ bins can always be granted as long as \eqref{eq:IV1} is satisfied.
\end{NewProof}

Theorem~\ref{thm:2} shows that fully loaded IFDMA systems can be achieved even without sorting the synchronous requests. In fact, the benefits of the min-small-change policy go far beyond this. Unlike the sort-first bin-filling policy, the min-small-change policy can be further applied in 1) asynchronous IFDMA systems where the requests come and go, as will be expounded in Section~\ref{sec:V}; and 2) systems wherein a subset of subcarriers are not available for use, e.g., the direct-conversion receivers in Section~\ref{sec:IVC} below.

\subsection{Subcarrier Allocation for Direct-Conversion Receiver}\label{sec:IVC}
For many communication systems, the RF signals are directly down-converted to the zero frequency (baseband) instead of the intermediate frequency (IF) for the simple design of direct-conversion receivers (DCR) \cite{DCR}. However, such systems typically incur a strong interfering signal at the DC because of the leakage of the local oscillator (LO) or the nonzero DC offset in the receiver's analog-to-digital converter (ADC). This is sidestepped in many systems, e.g., WiFi and LTE, by just not transmitting a subcarrier at the center of the band. Thus, it is desirable to avoid the use of signals containing DC components.

For the sort-first bin-filling policy, the null subcarrier at DC will be a problem. Considering the example in Table~\ref{tab:1}. When using DCR, the fourth subcarrier (the first bin) is unavailable, hence the request of user $B$ cannot be satisfied. On the other hand, for the min-small-change policy, avoiding the DC subcarrier is not a problem. Corollary~\ref{thm:3} below is a rephrase of the Theorem~\ref{thm:1} when the DC subcarrier is unavailable in DCR and with the use of the min-small-change policy.

\begin{cor}[Subcarrier allocation for DCR]\label{thm:3}
Let $a_n$ be the number of users requesting for $2^n$ subcarriers, and $M=2^m$. Consider direct-conversion receivers where the DC subcarrier is unavailable. Provided that the inequality in~\eqref{eq:IV4} below is satisfied, then we can use the bit-reversal subcarrier allocation scheme with the min-small-change bin-filling policy to assign subcarriers to users such that their IFDMA constraints are satisfied.
\begin{eqnarray}\label{eq:IV4}
\sum_{n=0}^{m}a_n 2^n \leq 2^m - 1.
\end{eqnarray}
\end{cor}
\begin{comm}
The RHS of~\eqref{eq:IV4} is due to the fact that we are giving up one subcarrier.
\end{comm}
\begin{NewProof}
We can treat the unavailable subcarrier as one that has already been allocated to a fictitious request of size $1$ that was presented to us first in the allocation process.
\end{NewProof}

To illustrate Corollary~\ref{thm:3}, consider the example of $M=8$ ($m=3$). At the beginning, instead of having a complete unoccupied fillable bin subset $[0,M-1]$, we have multiple smaller unoccupied fillable bin subsets. Suppose that the DC subcarrier is subcarrier $\{4\}$. After bit reversal, the corresponding bin to be avoided is bin $\{1\}$. Thus, the initial fillable bin subsets are $\{0\}$, $\{2,3\}$, $\{4,5,6,7\}$. That is, it is as if we have already granted bin $1$ to a fictitious request that asks for one subcarrier before we begin the real bin allocation.

For the original IFDMA system with the DC subcarrier, we have $2^{m-n}$ potential fillable bin subsets for a request that asks for $2^n$ subcarriers.  For an IFDMA system that avoids the DC subcarriers, the number of possible allocations is just one less than the above. That is, we have $2^{m-n}-1$ potential allocations for requests that ask for $2^n$ subcarriers. It is not possible to grant a request for $2^m$ subcarriers now; but this is the case even for an OFDM system that avoids the DC subcarrier. For requests that ask for $2^{m-1}$ subcarriers, we have one potential fillable bin subset in IFDMA that avoids the DC subcarrier; but even for the OFDMA system that avoids the DC subcarrier, one can only grant one (but not two) request that asks for $2^{m-1}$ subcarriers. Compared with non-DC OFDMA system, non-DC IFDMA has the same ``capacity performance'' in that the fully loaded capacity is decreased by $1$ only.

To conclude this section, we reiterate that, other than the limit that the request sizes has to be a divisor of $M$ (Constraint $2$), IFDMA has no limitations compared with OFDMA in terms of synchronous resource allocation. Even Constraint $2$ can be removed with our multi-stream IFDMA scheme. IFDMA/multi-stream IFDMA, on the other hand, have better PAPR performance than OFDMA.

%% file: SecV.tex
In asynchronous IFDMA systems, the users' requests arrive at the AP asynchronously. At any time, a request is granted only if there is an unoccupied fillable bin subset in the system whose size is equal to or larger than the size of the request (if the size is larger, the unoccupied fillable bin subset can be broken to multiple unoccupied smaller fillable bin subsets, one of which can be used to fulfill the request). Otherwise, the request will be rejected. Each accepted request has a holding time and it will leave the system upon the expiry of the holding time.

This section studies the subcarrier allocation problem in asynchronous IFDMA systems. Two issues of interest are as follows: 1) What is required in order for the IFDMA system to be {\it strictly nonblocking} (to be defined later)? 2) What is the performance gap between asynchronous IFDMA and OFDMA systems as far as the blocking probability is concerned?

To ease exposition, we denote the state of an IFDMA systems with $M$ subcarriers by the occupancies of $M$ bins $h_{M-1}h_{M-2}\cdots h_{1}h_{0}$. In particular, each binary symbol $h_i=\{0,1\}$ indicates whether the $i$-th bin is occupied: $h_i=0$ means it is vacant, and $h_i=1$ means it is occupied. Further, underlines below $h_i=1$ indicate grouping of the bins for a request. For example, consider the case of $M=4$. State $0~0~\underline{1~1}$ means the last two bins are allocated to a single request of size two, and state $0~0~\underline{1}~\underline{1}$ means the last two bins are allocated to two different requests of size one.

\subsection{Blocking Properties of Asynchronous IFDMA Systems}\label{sec:VA}
With the sort-first or the min-small-change policy, no request will be rejected in synchronous IFDMA systems as long as the total size of all requests does not exceed $M$. The reason is that the two bin-filling policies minimize fragmentation of unoccupied fillable bin subsets. For example, when $M=4$, $\underline{1}~\underline{1}~0~0$ is a legal state, but $\underline{1}~0~0~\underline{1}$ is illegal with the min-small-change policy when applied to synchronous IFDMA systems: in the first case, the two $0$'s represent an unoccupied fillable bin subset of size $2$, while in the second case, the two $0$'s represent two unoccupied fillable bin subsets of size $1$, which does not conform to the min-small-change policy.

On the other hand, even with the min-small-change bin-filling policy, $\underline{1}~0~0~\underline{1}$ is legal in asynchronous IFDMA systems because of departures of ongoing requests. For example, state $\underline{1}~0~\underline{1}~\underline{1}$ is legal because there could be three ongoing requests of size $1$. In an asynchronous IFDMA system, state $\underline{1}~0~\underline{1}~\underline{1}$ can evolve to state $\underline{1}~0~0~\underline{1}$ upon the departure of the request holding the third bin. If a new request of size $2$ then arrives, the new request cannot be granted even though there are two vacant bins in $\underline{1}~0~0~\underline{1}$ because the two vacant bins cannot form a fillable bin subset of size $2$.

For asynchronous IFDMA, if we allow the rearrangement of the occupied bins, there will also be no rejections in asynchronous IFDMA systems provided~\eqref{eq:IV1} is satisfied. For the same example in the above, a new request of size two can still be granted even when the system is in state $\underline{1}~0~0~\underline{1}$, because we can reassign bins to the existing requests so that the state becomes $\underline{1}~\underline{1}~0~0$ or $0~0~\underline{1}~\underline{1}$, creating an unoccupied fillable bin subset of two.

For concreteness, let us state the formal definitions of ``Rearrangeably Nonblocking'' and ``Strictly Nonblocking''~\cite{SwitchingBook}.

\begin{defi}[Rearrangeably Nonblocking]\label{def:4}
A system is said to be rearrangeably nonblocking if a new request can always be granted, after rearranging the assigned bins to the existing requests if necessary, provided that~\eqref{eq:IV1} is satisfied (i.e., the total size of the requests, including the new request, does not exceed $M$).
\end{defi}

\begin{prop}
Asynchronous IFDMA systems are rearrangeably nonblocking.
\end{prop}

\begin{NewProof}
Since we are allowed to reassign bins for the existing requests, we can rerun the synchronous resource allocation algorithm for all the requests (the existing requests plus the new request) as if they arrive synchronously.
\end{NewProof}

\begin{defi}[Strictly Nonblocking]\label{def:5}
A system is said to be strictly nonblocking if a new request can always be granted without rearranging the assigned bins to the existing requests, provided that~\eqref{eq:IV1} is satisfied (i.e., the total size of the requests, including the new request, does not exceed $M$).
\end{defi}

\begin{prop}
OFDMA systems are strictly nonblocking.
\end{prop}

\begin{NewProof}
This is obvious because a new request can be granted in OFDMA systems as long as there are enough vacant subcarriers, i.e., \eqref{eq:IV1} is satisfied.
\end{NewProof}

Asynchronous IFDMA poses a stringent requirement on the offered load if the system is to be strictly nonblocking. The RHS of \eqref{eq:IV1} must be relaxed to guarantee that there is no rejection/blocking of new requests.

\begin{defi}[Strictly Nonblocking under Offered Load Limit]\label{def:6}
A system is said to be strictly nonblocking with number of occupied bins limited to $K$ if a new request can always be granted, provided the total size of all requests (existing plus the new request) is no more than $K$.
\end{defi}

\begin{thm}[Strictly Nonblocking Property of Asynchronous IFDMA Systems under Offered Load Limit]\label{thm:6}
Suppose that $M=2^m$ is the total number of subcarriers, and there is a new request arriving at an asynchronous IFDMA system. Denote by $a_n$ the number of requests (including the new request) requiring $2^n$ subcarriers. The asynchronous IFDMA system is strictly nonblocking if and only if the following inequality is satisfied:
\begin{eqnarray}\label{eq:V1}
\sum_{n=0}^{m}a_n 2^n <
\left\{
\begin{array}{rcl}
2^{\frac{m}{2}+1}       &      & {\textup{if}~m~\textup{is even},}\\
3\times 2^{\frac{m-1}{2}}     &      & {\textup{if}~m~\textup{is odd}.}
\end{array} \right.
\end{eqnarray}
\end{thm}

\begin{NewProof}
Subcarrier occupancy can translate from bin occupancy through bit-reversal mapping. Thus, we can think in terms of bin occupancy.
We ask the following question. Suppose that there is a new request for $2^n$ subcarriers. What is the minimum number of bins occupied by existing users that can block the new request? For this new request, there are $M/2^n=2^{m-n}$ potential fillable bin subsets that can be used to fulfill it. However, if some bin in each and every of these potential fillable bin subsets is already occupied by an existing request, then none of them can be used to fulfill the new request. The minimum number of occupied bins to cause this situation is $2^{m-n}$, wherein one bin in each of the potential fillable bin subsets is occupied by an existing request of size $1$. If, on the other hand, the number of occupied bins is less than $2^{m-n}$,  then at least one of the potential fillable bin subsets is empty, and blocking cannot occur for the new request. Thus, to guarantee nonblocking property, the number of occupied bins plus the number of newly requested bins must be less than $2^{m-n}+2^n$. The RHS of \eqref{eq:V1} is found by $\min_{0\leq n\leq m}(2^{m-n}+2^n)$ (i.e., finding the worst-case $n$).
\end{NewProof}

Theorem~\ref{thm:6} indicates that, for asynchronous IFDMA systems, the price to pay for having the strictly nonblocking property is pretty steep and may not be worthwhile. Compared with \eqref{eq:IV1}, the total number of requested subcarriers allowed at any one time has to be severely limited in order to have the strictly nonblocking property as in \eqref{eq:V1}.

\textbf{Multi-stream IFDMA} -- A multi-stream IFDMA scheme can enable strictly nonblocking IFDMA systems with full loading (i.e., a multi-stream IFDMA system is strictly nonblocking in the sense of Definition~\ref{def:5}).

As stated in Definition~\ref{def:2}, a multi-stream IFDMA scheme allows the user to construct multiple IFDMA streams, as opposed to only one stream in traditional IFDMA systems. Given a new request of size $N$, the AP can partition the request to multiple sub-requests, and the size of each sub-request is a divisor of $M$. In this context, the problem of finding an unoccupied fillable bin subset of size $N$ is transformed to a problem of finding multiple unoccupied fillable bin subsets of smaller size. In the extreme case, the AP can partition the request to $N$ sub-requests, each of size one.

\begin{prop}
Multi-stream IFDMA systems are strictly nonblocking.
\end{prop}

\begin{NewProof}
If the number of total requested subcarriers does not exceed $M$, a new request can always be partitioned to multiple sub-requests so that multiple empty fillable bin subsets can be used to fulfill the request ¨C- in the extreme case, we partition the incoming request to multiple sub-requests of size one.
\end{NewProof}

In addition to the strictly nonblocking property, multi-stream IFDMA also enables fine-grained asynchronous subcarrier allocation. The size of requests is restricted to a divisor of $M$ in Theorem~\ref{thm:6}. This restriction can be removed with the multi-stream scheme. Each user can request an arbitrary number of subcarriers, and the multi-stream IFDMA systems are strictly nonblocking as long as the total number of requested subcarriers is no more than $M$.

\subsection{Blocking Probability of Asynchronous IFDMA Systems}\label{sec:VB}
Without the multi-stream scheme, the strictly nonblocking property is expensive to achieve for asynchronous IFDMA systems. Let $M = 1024$ ($m = 10$), for example. To have a strictly nonblocking IFDMA system, the load needs to be limited to $2^{\frac{m}{2}+1}=64$ subcarriers according to Theorem~\ref{thm:6}. On the other hand, OFDMA systems are strictly nonblocking even with full loading of $1024$ subcarriers.

However, given a certain amount of blocking tolerance (i.e., we give up making the system strictly nonblocking by allowing blocking as long as the blocking probability is not excessive), how does an IFDMA system perform when benchmarked against an OFDMA system in the asynchronous scenario? This subsection answers this question.

In asynchronous IFDMA systems, the bin-filling policy is very important in that it determines the probability that a new request gets blocked. We will consider two bin-filling polices for asynchronous IFDMA: the random policy considered in \cite{Allocation1,Allocation2}, and our min-small-change policy proposed in Section~\ref{sec:IV}.

\noindent\textbf{The Markov Chain} -- Analytically, the state transitions in asynchronous IFDMA systems can be modeled as a Markov Chain (MC) \cite{StochasticBook} for a given request arrival model (e.g., Poisson arrival), a given holding-time model (e.g., exponential holding time), and a given bin-filling policy.

\begin{figure}[t]
  \centering
  \includegraphics[width=0.7\columnwidth]{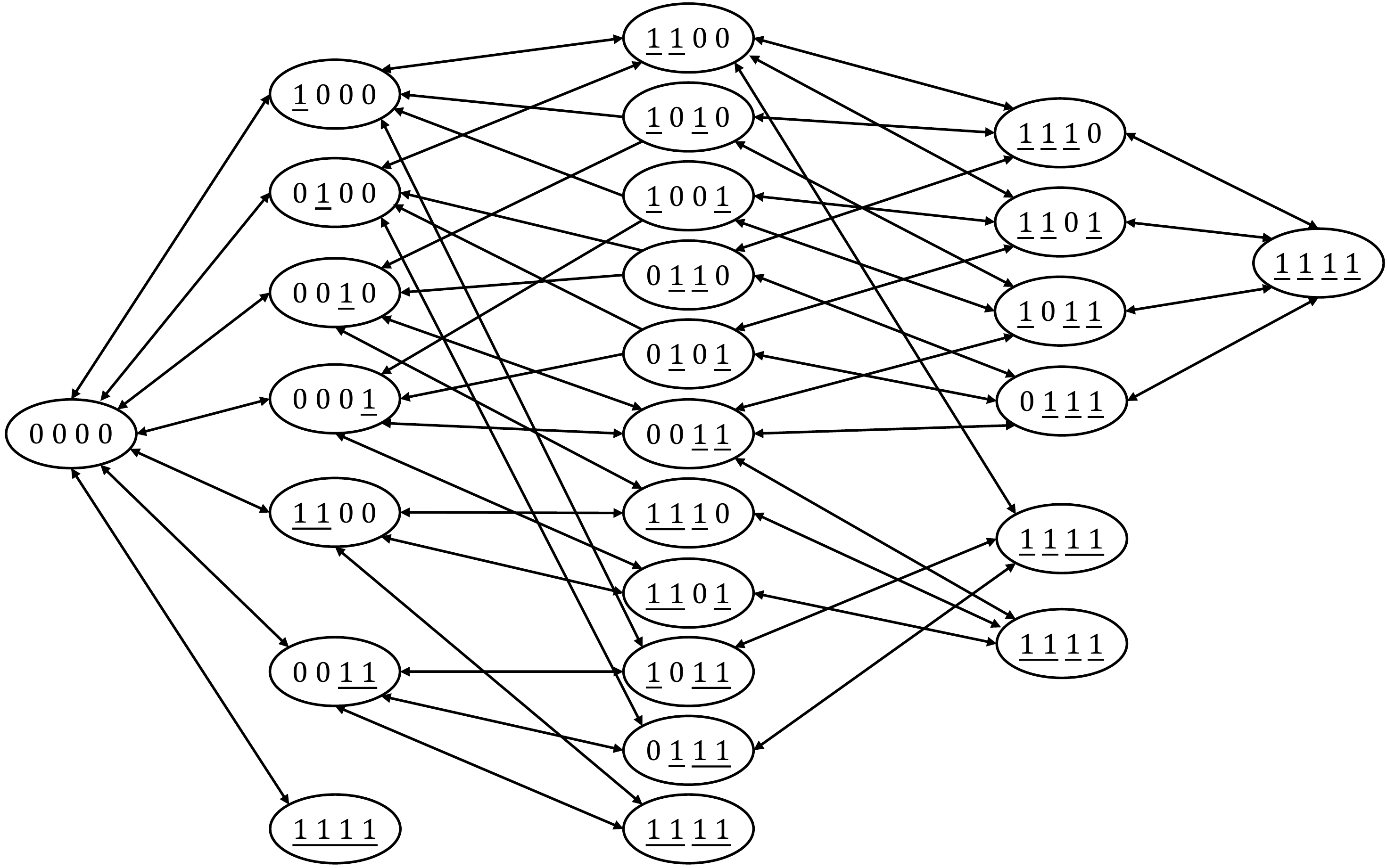}\\
  \caption{An example of the MC associated with the state transitions of asynchronous IFDMA systems. The total number of subcarriers $M = 4$, and we use the min-small-change policy to fill the bins. If the random bin-filling policy is used, all the edges in the MC are bidirectional.}
\label{fig:5}
\end{figure}

The complete state transitions of the MC when $M = 4$ is shown in Fig.~\ref{fig:5}, where the min-small-change policy is used to fill the bins (we omit the state-transition probabilities to avoid cluttering the picture). As can be seen, there are $26$ states. A bidirectional edge between two states means the transition can be from left to right (upon a request arrival), and also from right to left (upon a request departure). A unidirectional links means the state can only transit in one direction, in particular, from right to left. For instance, suppose the system is in state $\underline{1}~0~0~0$. Upon an arrival of request of size one, the state can become $\underline{1}~\underline{1}~0~0$, but not $\underline{1}~0~\underline{1}~0$ or $\underline{1}~0~0~\underline{1}$ according to the min-small-change policy. On the other hand, all three states $\underline{1}~\underline{1}~0~0$, $\underline{1}~0~\underline{1}~0$ and $\underline{1}~0~0~\underline{1}$ can potentially evolve to state $\underline{1}~0~0~0$ upon a departure of a request of size one. Thus, the edge between states $\underline{1}~0~0~0$ and $\underline{1}~\underline{1}~0~0$ is bidirectional, while the edge between $\underline{1}~0~0~0$ and $\underline{1}~0~\underline{1}~0$ (or $\underline{1}~0~0~\underline{1}$) is unidirectional.

To derive the blocking probability of the IFDMA system, we have to compute the steady-state distribution of the associated MC. However, the number of the states in the MC grows at an extremely fast rate (even faster than exponential) with the increase of $M$. As indicated by Proposition~\ref{thm:8}. The MC becomes analytically intractable very quickly.

\begin{prop}[Number of states in the MC]\label{thm:8}
Let $M=2^m$ be the total number of subcarriers, and $f(m)$ be the number of states in the MC associated with the asynchronous IFDMA systems. We have
\begin{eqnarray}\label{eq:V2}
f(m)=f^2(m-1)+1\geq 2^{2^m}.
\end{eqnarray}
\end{prop}

\begin{NewProof}
See Appendix~\ref{sec:AppB}.
\end{NewProof}

\begin{figure}[t]
  \centering
  \includegraphics[width=0.7\columnwidth]{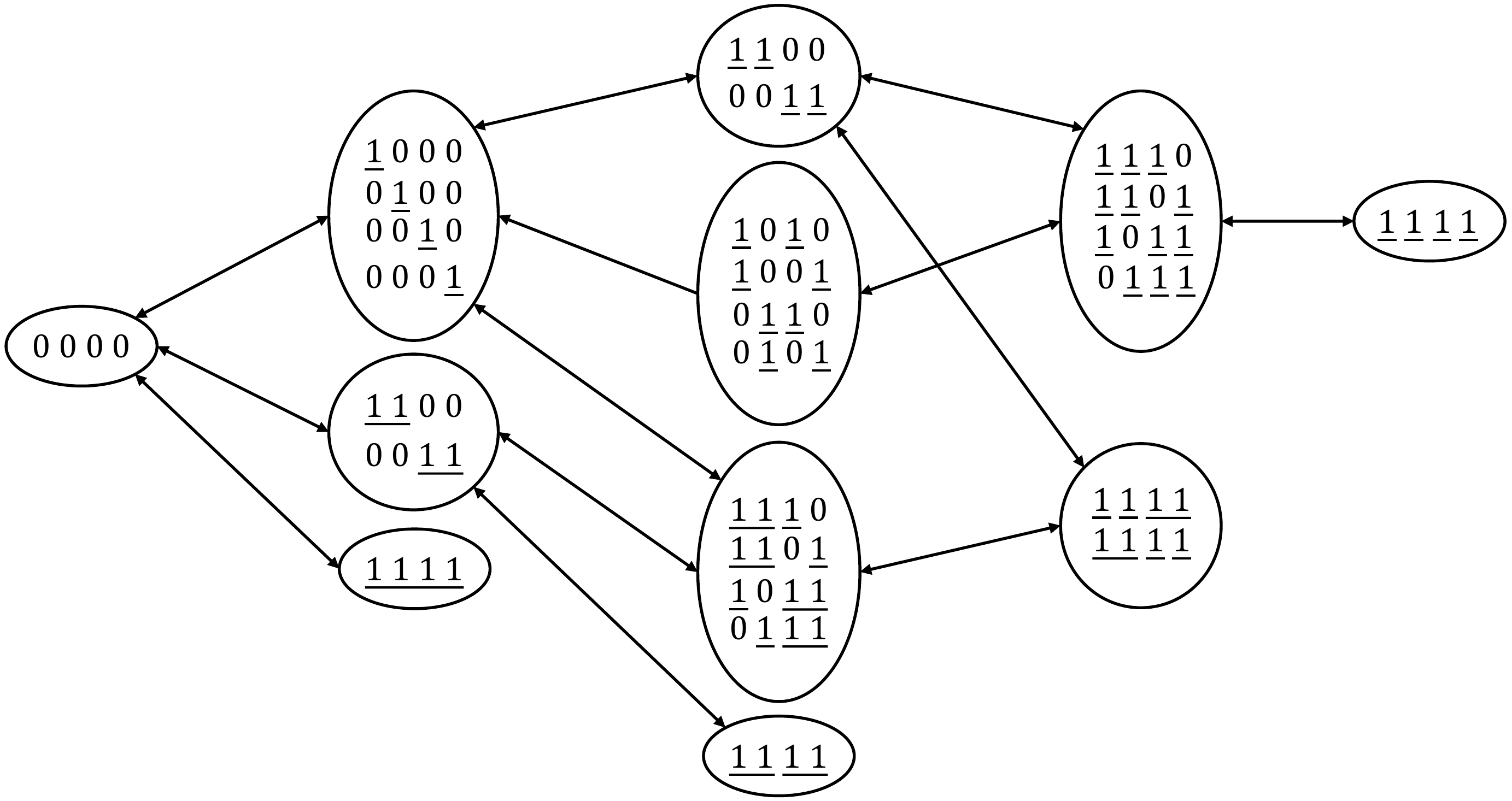}\\
  \caption{An example of the consolidated MC associated with the state transitions of asynchronous IFDMA systems. The total number of subcarriers $M = 4$, and we use the min-small-change policy to fill the bins.}
\label{fig:6}
\end{figure}

A possible way to reduce the number of states in the MC is to merge the states that have the same blocking probabilities into a super state. For example, the states $\underline{1}~0~0~0$, $0~\underline{1}~0~0$, $0~0~\underline{1}~0$, and $0~0~0~\underline{1}$ can be merged together as a super state because they are equivalent in terms of blocking probability. The MC then becomes a consolidated MC, in which the number of super states is much less than the number of fine states in the original MC. As an example, the fine-state MC in Fig.~\ref{fig:5} can be transformed to the consolidated MC in Fig.~\ref{fig:6}, where the number of super states is $11$, while the number of fine states is $26$.

\begin{prop}[Number of super states in the consolidated MC]\label{thm:9}
Let $M=2^m$, and $g(m)$ be the number of super states in the consolidated MC associated with the subcarrier allocation in asynchronous IFDMA systems. We have
\begin{eqnarray}\label{eq:V3}
g(m)=\frac{1}{2}g^2(m-1)+\frac{1}{2}g(m-1)+1\geq 2^{2^{m-1}+1}.
\end{eqnarray}
\end{prop}

\begin{NewProof}
See Appendix~\ref{sec:AppC}.
\end{NewProof}

Proposition~\ref{thm:9} shows that the number of super states is still too huge to handle. In view of the analytical intractability of the problem, we resort to simulation analysis to investigate the blocking probability of asynchronous IFDMA systems.

\noindent\textbf{Monte-Carlo Simulations} -- We simulate a multiple access system with one AP and multiple users. The total number of subcarriers is $M=2^m=1024$, and the users can request $N=2^n$, $n=0,1,2,\cdots,m$ subcarriers. The system is either IFDMA or OFDMA.

For requests of different sizes, we assume their arrivals follow independent Poisson process. In particular, the arrival rate (the number of arrivals per unit time) of a request is inversely proportional to its size, i.e., for a request of size $2^n$, the arrival rate $\lambda_n=\frac{1}{2^n}\lambda$ where $\lambda$ is a constant. Further, we assume the holding time of a request (i.e., the duration that the user stays in the system) follows exponential distribution with mean $\mu$. Without loss of generality, we assume $\mu=1$ for all requests.

For this system, the offered load (normalized by the total number of subcarriers) is given by
\begin{eqnarray}\label{eq:V4}
G=\frac{1}{2^m}\sum_{n=0}^{m}\frac{2^n\lambda_n}{\mu}=(m+1)2^{-m}\lambda.
\end{eqnarray}

The blocking probability can be computed by
\begin{eqnarray}\label{eq:V5}
P_B=\frac{\sum_{n=0}^{m}2^n r_B(n)}{\sum_{n=0}^{m}2^n r(n)}
\end{eqnarray}
where $r(n)$ is the number of arrived requests of size $2^n$, and $r_B(n)$ is the number of requests of size $2^n$ that gets blocked. The normalized throughput of the system $S=1-P_B$.

Under random arrivals, a request is sure to be blocked in the case of instantaneous overload, wherein there are not enough unoccupied subcarriers left to fulfil the number of subcarriers requested. We note that even for a strictly nonblocking system as defined in Definition~\ref{def:5}, blocking due to instantaneous overload can still occur under the random request arrival model.

As summarized in Fig.~\ref{fig:4}, instantaneous overload is the only cause for blocking in OFDMA. On the other hand, in IFDMA, a request can be blocked due to either instantaneous overload or fragmentation of bins, in which case the number of vacant bins is larger than the size of the request, but there is no unoccupied fillable bin subset available to fulfill the request.

In this context, we further define the ``the probability of blocking due to fragmentation'' to characterize the proportion of the ``rejected load due to fragmentation'' in the offered load:
\begin{eqnarray}\label{eq:V6}
P_f=\frac{\sum_{n=0}^{m}2^n r_f(n)}{\sum_{n=0}^{m}2^n r(n)}
\end{eqnarray}
where $r_f(n)$ is the number of requests of size $2^n$ that gets blocked when the system backlog (i.e., the number of occupied subcarriers) is less than or equal to $2^m-2^n$ (i.e., blocking only due to the fragmentation of bins). Note that $P_f=0$ in OFDMA, because a request of size $2^n$ can be granted as long as the system backlog is less than or equal to $2^m-2^n$.

\begin{figure}[t]
  \centering
  \includegraphics[width=0.6\columnwidth]{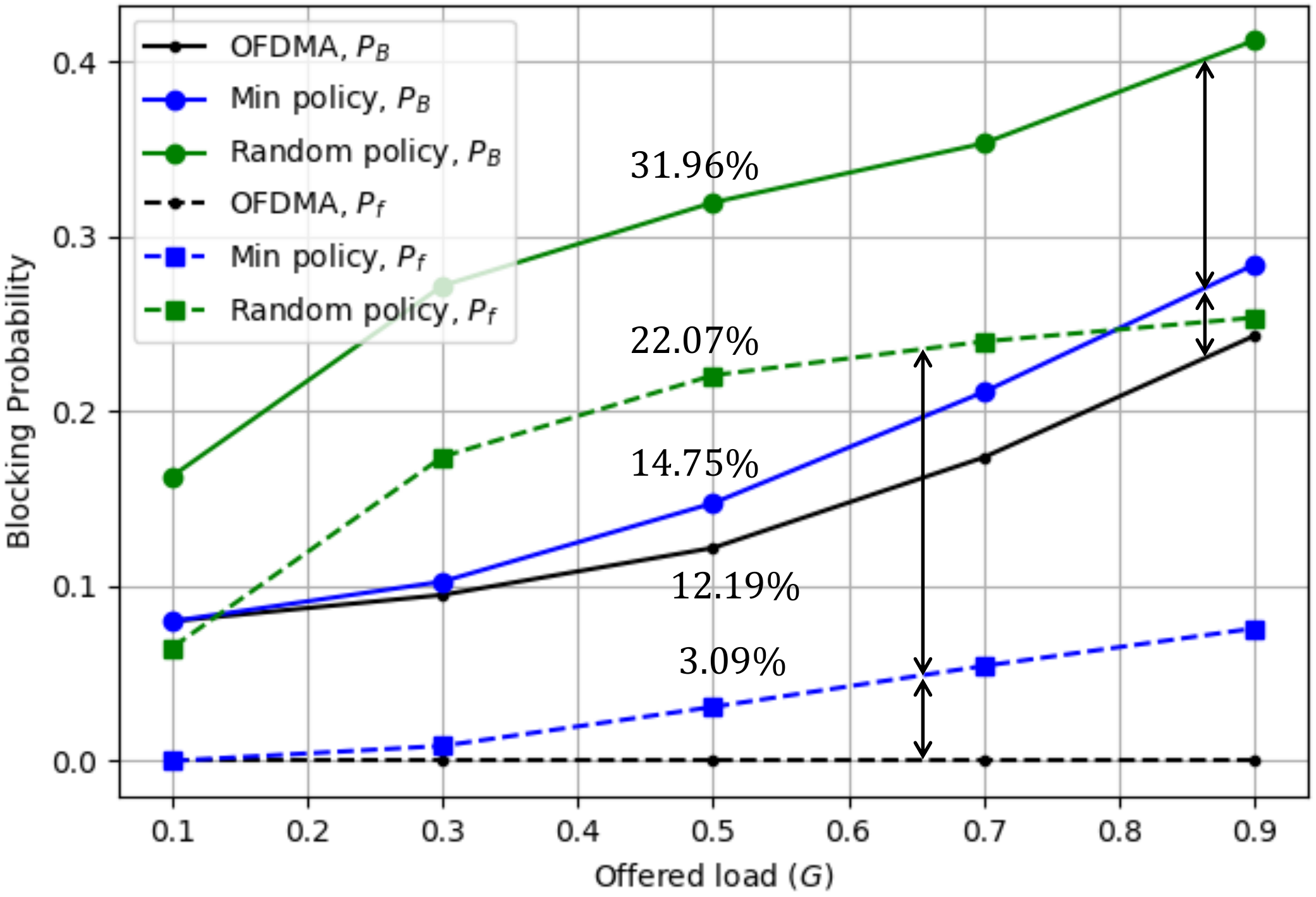}\\
  \caption{The blocking probabilities versus the offered load of OFDMA and IFDMA systems.}
\label{fig:7}
\end{figure}

Fig.~\ref{fig:7} presents the blocking probabilities ($P_B$ and $P_f$) versus the offered load to the system. Three schemes are simulated: OFDMA, IFDMA with random bin-filling policy, and IFDMA with the min-small-change bin-filling policy. They are abbreviated as ``OFDMA'', ``Random policy'', and ``Min policy'' in the figure, respectively.

Let us first compare the two bin-filling polices in IFDMA systems. As shown, the min-small-change policy performs significantly better than the random policy. At a moderate offered load $0.5$, the min-small-change policy reduces the blocking probability $P_B$ by $17.21\%$.

The performance of $P_f$ is even more revealing to showcase the importance of a good bin-filling policy. For the min-small-change policy, $3.09\%$ of the offered load are rejected due to the fragmentation of bins. On the other hand, for the random policy, $22.07\%$ of the offered load are rejected due to the fragmentation of bins. The min-small-change policy reduces the blocking probability due to the fragmentation of bins by $18.98\%$.

Then, we benchmark the blocking probabilities of IFDMA systems (with the min-small-change policy) against the OFDM systems. As can be seen, as expected, IFDMA systems has higher blocking probability than the OFDMA systems, but the gap is not large. When the offered load is $0.5$, the performance gap is only $2.56\%$ in terms of $P_B$, and $3.09\%$ in terms of $P_f$.

In the simulation of Fig.~\ref{fig:7}, requests of different sizes (from size $1$ to size $M$) are mixed together and fed into the systems. Additional simulations are performed by us to investigate the blocking probabilities of requests of different sizes. We found that, unsurprisingly, the requests with larger size have higher blocking probabilities, in particular, the requests of size $M$ are mostly blocked (again this is obvious because even there is one occupied bin only, a request of size $M$ will be blocked). This is the case for both IFDMA and OFDMA.

Whether for IFDMA or OFDMA systems, it is not a good idea to support a mix of large and small requests at the same time. This is because large requests will be blocked most of the time, effectively eliminating the use of the system to fulfill large requests.

In this light, we advocate not mixing requests of small and large sizes into the same system. To verify the benefits of this, we repeat the simulations in Fig.~\ref{fig:7}, but change the composition of the traffic fed into the system. Specifically, given the same offered load to the system, the sizes of the request in Fig.~\ref{fig:7} can be $2^n$ where $n=0,1,\cdots,m$; however in the new simulation, we limit the size of the requests to $2^n$ where $n=0,1,\cdots,m/2$.\footnote{The proof of Theorem~\ref{thm:6} indicates that the condition for strictly nonblocking is the most stringent when $n=m/2$. That is, $\arg\min_{0\geq n\geq m}(2^{m-n}+2^n)$ is in the ballpark of $m/2$. Thus, we limit the size of the request to be no more than $2^{m/2}$ so that we are still considering the ``worst-case'' of blocking due to fragmentation. Essentially, we are reducing blocking due to instantaneous overload while retaining the worst cases of blocking due to fragmentation.}

\begin{figure}[t]
  \centering
  \includegraphics[width=0.6\columnwidth]{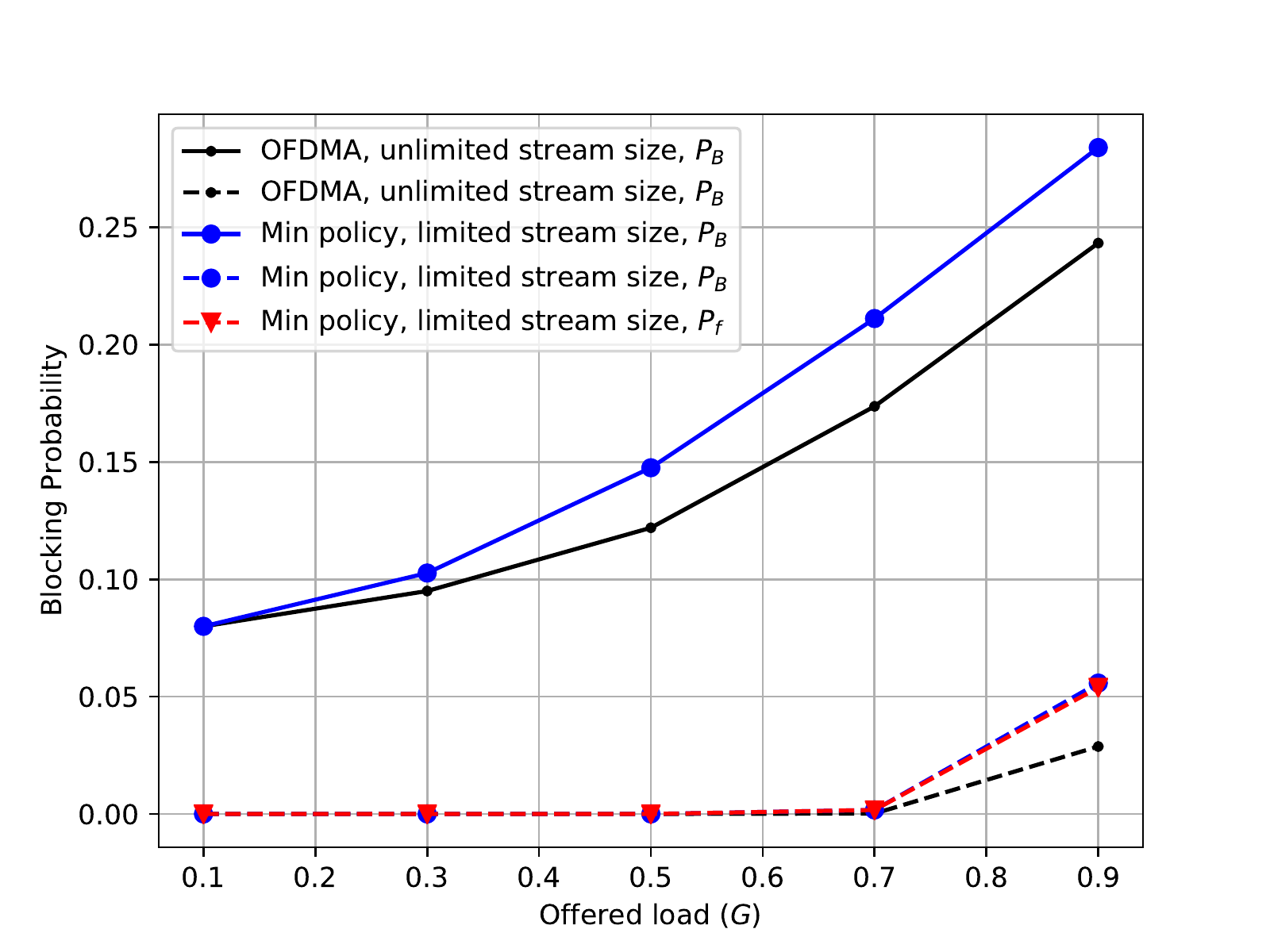}\\
  \caption{The blocking probabilities of OFDMA and IFDMA versus the offered load to the system. The solid curves are copied from Fig.~\ref{fig:7}, and the dashed curves are the results of limiting the stream size.}
\label{fig:8}
\end{figure}

The simulation results are presented in Fig.~\ref{fig:8}. As shown, for the same offered load, the blocking probabilities $P_B$ of both OFDMA and IFDMA systems are reduced by much under the new traffic model where the mixing of small and large streams is prevented.
In particular, for IFDMA, almost all the blocking are caused by the fragmentation of bins at a heavy load $0.9$. This matches with our traffic design that the worst case of blocking due to fragmentation is retained, only the blocking due to instantaneous overload is largely reduced.

Furthermore, the performance gaps between asynchronous OFDMA and IFDMA systems are negligible under light and moderate offered load. This is reasonable because the span of the stream size is limited.
We note in passing also that, to the extent that all requests are of the same size (i.e., $2^n$ for a fixed $n$), then there is no difference between IFDMA and OFDMA as far as blocking probability is concerned.

%% file: AppendixA.tex
The foundation of many of our results in Section~\ref{sec:IV} and \ref{sec:V} is the bit-reversal subcarrier allocation scheme, where we assume the total number of subcarriers $M$ is a power of $2$. For general composite $M$, the bit-reversal procedure can be generalized to a digit-reversal procedure. With digit reversal, the results for the power-of-$2$ $M$ case can then be generalized to those for general composite $M$.

Let us consider a synchronous IFDMA system to elaborate the idea of digit-reversal subcarrier allocation. Let $M$ be the total number of subcarriers in the system, where $M$ is a composite number. A prime factorization of $M$ is an operation of factoring $M$ into a product of $T$ primes in the order of $p_{T-1}$, $p_{T-2}$, $\cdots$, $p_{1}$, $p_{0}$, i.e., $M=p_{T-1}\times p_{T-2}\times \cdots\times p_{1}\times p_{0}$. The prime factorization of $M$ is not unique in that the order of the primes can vary. Power-of-$2$ $M$ is a special case, however, in that all prime factors are $2$, and therefore the order does not matter and the prime factorization is unique.

When the prime factorization of $M$ is not unique, we can only choose one of the possible prime factorizations to serve as the architectural bedrock for our digit-reversal subcarrier-allocation algorithm and the transceiver designs (see our companion paper \cite{tech2}). Intermixing of different factorizations by different users (i.e., different users assume different factorizations) will lead to inconsistencies in the subcarrier allocation, as well as the multiplexing and demultiplexing processes in the transceiver design.

Given a decomposition $M=p_{T-1}\times p_{T-2}\times \cdots\times p_{1}\times p_{0}$, the IFDMA system can only support requests of sizes $\{1,p_0,p_0p_1,p_0p_1p_2,\cdots,p_0p_1...p_{T-1}\}$. For example, there are three possible factorizations for $M = 12$: $12=3\times 2\times 2$, $12=2\times 3\times 2$, and $12=2\times 2\times 3$. The three corresponding sets of allowed request sizes are $\{1, 2, 4, 12\}$, $\{1, 2, 6, 12\}$, and $\{1, 3, 6, 12\}$.

This resource allocation constraint applies to single-stream IFDMA system only, and for multi-stream IFDMA, requests of any size can be satisfied by allocating multiple streams of IFDMA to a request. For example, a request of size $6$ can be satisfied by an IFDMA stream of size $2$ plus an IFDMA stream of size $4$ if the factorization that leads to allowed stream sizes of $\{1, 2, 4, 12\}$ is adopted. On the other hand, the same request can be satisfied by an IFDMA stream size of $6$ if the factorization that leads to allowed stream sizes of $\{1,2,6,12\}$ is adopted instead.

We note, however, that even for multi-stream IFDMA, the transceiver design still depends on the factorization adopted, and that for consistency of the system, all users must assume the same factorization. Although the sizes of requests are not limited, how to construct multiple streams to satisfy a request does depend on the factorization adopted, and hence the transceiver design~\cite{tech2}.

The rest of this subsection focuses on single-stream IFDMA. To illustrate, we consider the generalization of the sort-first bin-filling policy

\vspace{0.4cm}
\noindent \textbf{Algo 2: The digit-reversal subcarrier allocation scheme.}

\noindent Suppose that we have $M=\prod_{t=0}^{T-1}p_t$ bins labelled with $k=0,1,\cdots,M-1$. In particular, an index $k$ can be expressed in the digit form as $d_{T-1}d_{T-2}...d_{1}d_{0}$, where $d_t=0,1,2,...,p_t-1$.
\begin{itemize}
\item Sorting. We sort the users according to the number of requested subcarriers in descending order.
\item Bin Allocation. We allocate bins to the sorted users in the order of bin $0$ to bin $M-1$. That is, each user who requests $N_i$ subcarriers is allocated the next $N_i$ unallocated bins. Users with larger request size is considered first.
\item Bit-reversal mapping. We perform a bit-reversal mapping from bins to subcarriers. That is, if a bin with binary index $d_{T-1}d_{T-2}...d_{1}d_{0}$ is allocated to a user, then the subcarrier with binary index $d_{0}d_{1}... d_{T-2}d_{T-1}$ will be allocated to this user.
\end{itemize}

\vspace{0.4cm}
Table~\ref{tab:2} gives an example of the digit-reversal subcarrier allocation scheme. In this example, we assume $M=12$, and the prime factorization of $M$ is fixed to $12=2\times 2\times 3$ (i.e., $p_0=3$, $p_1=2$, and $p_2=2$). In this system, the request sizes can be $\{1, 3, 6, 12\}$. Suppose that there are three users $A$, $B$, and $C$ requesting $3$, $1$, and $6$ subcarriers, respectively. To use the digit-reversal subcarrier allocation scheme, we first sort the users according to their requests. The sorted users are $C$ ($6$ subcarriers), $A$ ($3$ subcarriers) and $B$ ($1$ subcarrier). Following the second step, we allocate bins $\{0, 1, 2, 3, 4, 5\}$ to user $C$, bins $\{6, 7, 8\}$ to user $A$, and then bins $\{9\}$ to user $B$. After digit-reversal mapping, users $A$, $B$ and $C$ finally get subcarriers indexed by $\{1, 5, 9\}$, $\{3\}$, and $\{0,4,8,2,6,10\}$, respectively.

\begin{table}[t]
\caption{An example of the digit-reversal subcarrier allocation.}
\center
\begin{tabular}{cccc}
\toprule
\rowcolor[HTML]{EFEFEF}
\textbf{Bins}          & \textbf{Digit index of bins} & \textbf{Digit-reversal} & \textbf{Subcarriers} \\
\midrule
0 & $000$ & $000$ & $0$ \\
1 & $001$ & $100$ & $4$ \\
2 & $002$ & $200$ & $8$ \\
3 & $010$ & $010$ & $2$ \\
4 & $011$ & $110$ & $6$ \\
5 & $012$ & $210$ & $10$ \\
6 & $100$ & $001$ & $1$ \\
7 & $101$ & $101$ & $5$ \\
8 & $102$ & $201$ & $9$ \\
9 & $110$ & $011$ & $3$ \\
10 & $111$ & $111$ & $7$ \\
11 & $112$ & $211$ & $11$ \\
\bottomrule
\end{tabular}
\label{tab:2}
\end{table}

When operated with the digit-reversal subcarrier allocation scheme, we have similar claims to Theorem~\ref{thm:1} that full loading is possible for synchronous IFDMA systems with a composite $M$.

\begin{cor}[Full loading is possible for Composite-$M$ IFDMA]\label{thm:10}
In a synchronous IFDMA system with $M$ subcarriers where $M$ is a composite number, provided that the total number of requested subcarriers does not exceed $M$, we can use the digit-reversal subcarrier allocation scheme, with either the sort-first or the min-small-change bin-filling policy, to assign subcarriers to users such that the evenly-spaced-subcarrier constraint is satisfied.
\end{cor}

\begin{NewProof}
The general proof follows the proof of Theorem~\ref{thm:1} by induction. First, it can be easily verified that corollary~\ref{thm:10} is valid for $M=2$ or $3$. We then prove that if the theorem is valid for any $M=p_{T-1}\times p_{T-2}\times \cdots\times p_{1}\times p_{0}$, it is also valid for $M'=p_TM=p_T\times p_{T-1}\times p_{T-2}\times \cdots\times p_{1}\times p_{0}$.

Consider bin allocation associated with the second of the digit-reversal allocation algorithm. There are two possible cases for the users' requests.
Case $i$: There is a single user requesting $M'$ subcarriers. The theorem is trivially true for this case.
Case $ii$: No user requests $M'$ subcarriers. For this case, we divide the $M'$ bins into $pT$ fillable bin subsets $[iM,iM+M-1]$, $i=0,1,2,...,p_T-1$. Since there is no request with size more than $M$, the bins allocated to each user must all belong to one of the $p_T$ subsets according to the second step in Algo. 2. For each fillable bin subset, if we were to consider the subcarrier allocation problem with $M$ subcarriers, the spacing between subcarriers allocated to a request of size $N$ would be $M/N$. When we migrate to the subcarrier allocation problem with $M'$ subcarriers, it can be shown that the new spacing is $p_TM/N=M'/N$, thus satisfying the new IFDMA constraint.
\end{NewProof}

Besides the sort-first bin-filling policy presented in Algo. 2, it is straightforward that the min-small-change policy can also be used to achieve fully loaded IFDMA systems.

%% file: AppendixB.tex
\begin{NewProof}
To analyze the number of the states in the MC, let us represent the state of the system as a binary tree.

Initially, all bins are free to be used, and the system is in state $0~0~...~0~0$. The corresponding tree is simply a root node, i.e., an empty fillable bin subset of size $M=2^m$. Suppose, for example, that a request of size $2^{m-2}$ arrives, the state then becomes $\underline{1~...~1}~0~0~...~0$ ($2^{m-2}$ ones and $3\times 2^{m-2}$ zeros). The corresponding tree is shown in Fig.~\ref{fig:9}, where we use the black nodes to represent the occupied fillable bin subsets, the white nodes to represent the unoccupied fillable bin subsets, and the green nodes to represent the partially occupied fillable bin subset (a fillable bin subset that is partitioned into two fillable bin subsets, one of which is empty, and the other of which is either occupied or partially occupied).

\begin{figure}[t]
  \centering
  \includegraphics[width=0.3\columnwidth]{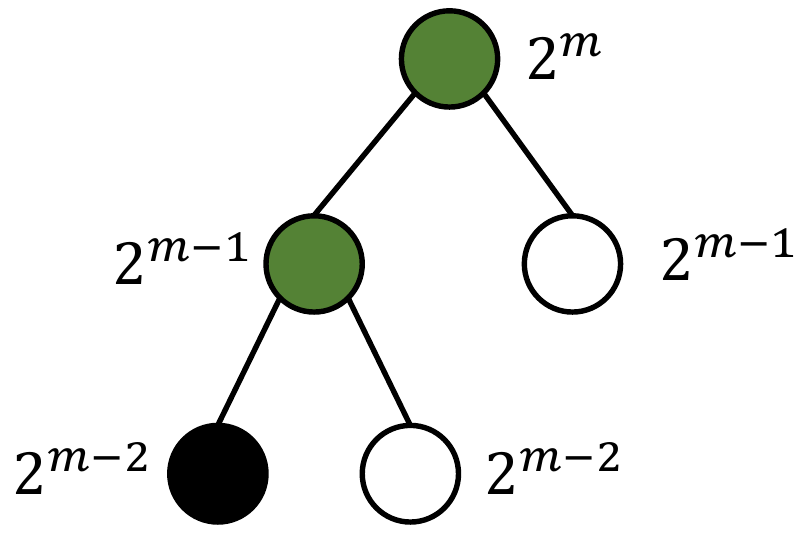}\\
  \caption{The state of the MC as a tree structure.}
\label{fig:9}
\end{figure}

One state of the system corresponds to one form of such a binary tree. We could count the total number of tree structures to get the total number of possible states $f(m)$.

Case I: We consider the single-node trees. There are two such trees: a single black root (a single request of size $M$ occupies all bins), and a single white root (all bins are free to be used).

Case II: We consider the trees that have nodes beyond the root node. The root node has two child nodes which form a left subtree and a right subtree. Notice that each of the left and right subtrees has $f(m-1)$ possible forms. Excluding the case that both left and right subtrees are a single white node (i.e., both are free to be used, but this is illegal this case corresponds to one of the subcases of case I, which has been taken into account already), we have $f^2(m-1)-1$ different tree structures in this case.

Overall, the number of tree structures, i.e., the number of possible states, is $f(m)=f^2(m-1)+1$. From $f(0)=2$, we can then show that $f(m)\geq 2^{2^m}$.
\end{NewProof}

%% file: AppendixC.tex
A super state groups all the fine states that have the same blocking probabilities. If we use the tree interpretation, the trees correspond to these states are ``isomorphic'' to each other.

\begin{figure}[t]
  \centering
  \includegraphics[width=0.55\columnwidth]{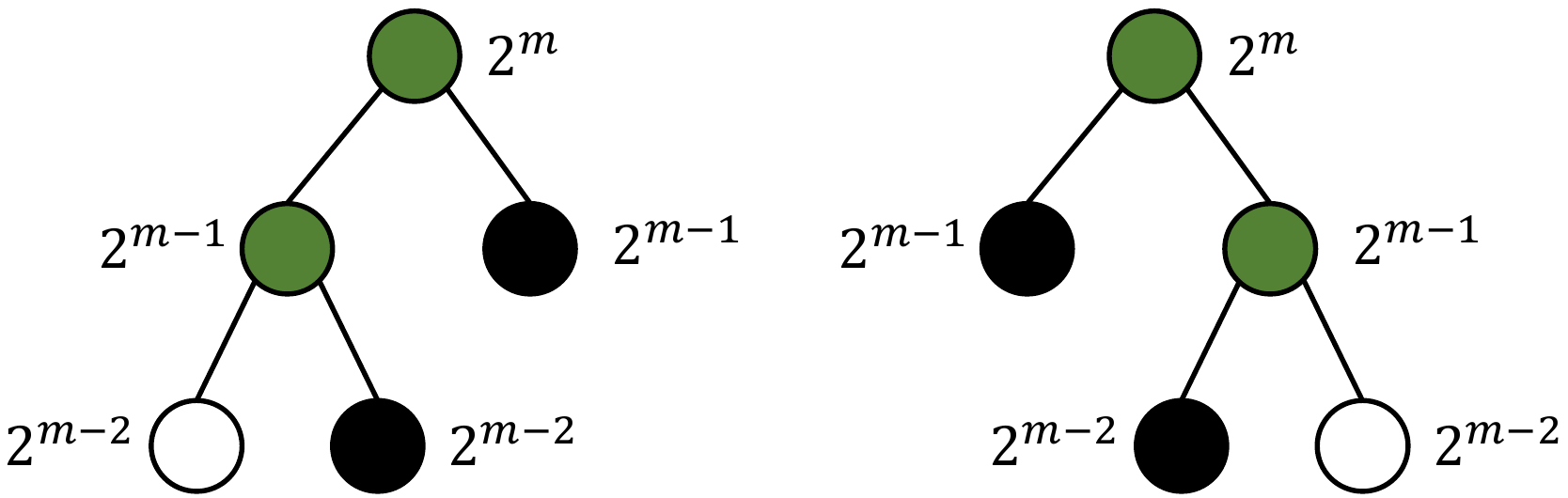}\\
  \caption{Two isomorphic trees.}
\label{fig:10}
\end{figure}

Two non-identical trees are isomorphic if by swapping the left and right branches of a series of subtrees embedded within one of the two trees (could be subtrees at different level of the tree), we can obtain the other tree. Two isomorphic trees are shown in Fig.~\ref{fig:10}. Let $m = 2$, the represented states of the two trees in Fig.~\ref{fig:10} are $0~\underline{1}~\underline{1~1}$ and $\underline{1~1}~\underline{1}~0$. These two states can be grouped together in the same super state, as in Fig.~\ref{fig:6}.

To compute the number of super states in the consolidated MC, $g(m)$, we only need to count the number of non-isomorphic trees.

Case I: We consider the single-node non-isomorphic trees. As in Proposition~\ref{thm:8}, there are two such trees: a single black root (a single request of size $M$ occupies all bins), and a single white root (all bins are free to be used).

Case II: We consider the trees that have nodes beyond the root node. The root node has two child nodes which form a left subtree and a right subtree, denoted by $T_l$ and $T_r$, respectively. The tree can then be represented by a duple $(T_l,T_r)$. Each of $T_l$ and $T_r$ has $g(m-1)$ non-isomorphic forms. The number of possible combinations of the subtrees to form the overall tree is $g^2(m-1)$, but some of the combinations are isomorphic to other combinations although there is no isomorphism within each subtree.

A tree $(T_l,T_r)$ is isomorphic to $(T_r,T_l)$, because if we switch the left subtree and the right subtree, the resulting tree is isomorphic to the original one. These isomorphic trees are double-counted in $g^2(m-1)$. In other words, if we take a form of the tree $(T_l,T_r)$ as an entry of a $g(m-1)\times g(m-1)$ matrix, then this matrix is symmetric, and the number of non-isomorphic forms is the number of upper-diagonal entries including the diagonal entries. This gives us $\frac{1}{2}g^2(m-1)+\frac{1}{2}g(m-1)$ non-isomorphic forms. Further, we have to exclude the case where both $T_l$ and $T_r$ are a single white root (this is a particular diagonal entry in the matrix).

Overall, the number of non-isomorphic tree structures, i.e., the number of super states in the consolidated MC, is $g(m)=2+\frac{1}{2}g^2(m-1)+\frac{1}{2}g(m-1)-1=\frac{1}{2}g^2(m-1)+\frac{1}{2}g(m-1)+1$. From $g(1)=2^2$, we can then show that $g(m)\geq 2^{2^{m-1}+1}$ for $m\geq 1$.